\documentclass{pasa}%

\newcommand\hi{\mbox{\sc Hi}}
\newcommand\msun{M$_{\odot}$}

\jid{PASA}
\doi{10.1017/pas.\the\year.xxx}
\jyear{\the\year}

\usepackage{aas_macros}
\usepackage{hyperref} 
\hypersetup{colorlinks,citecolor=blue,linkcolor=blue,urlcolor=blue}
\usepackage{amsmath}
\usepackage{mathrsfs}
\usepackage{graphicx}
\usepackage{enumitem}
\title[Pilot GASKAP-HI Observations of the SMC]{GASKAP-HI Pilot Survey Science I: ASKAP Zoom Observations of \hi~Emission in the Small Magellanic Cloud}
\author[Pingel et al.]{N.~M.~Pingel$^{1}$, J.~Dempsey$^{1,2}$, N.~M.~McClure-Griffiths$^1$, J.~M.~Dickey$^3$, K.~E.~Jameson$^4$, H.~Arce$^5$, G.~Anglada$^6$, J.~Bland-Hawthorn$^7$, S.~L.~Breen$^8$, F.~Buckland-Willis$^9$, S.~E.~Clark$^{10, 11}$, J.~R.~Dawson$^{12,13}$, H.~D\'{e}nes$^{14}$, E.~M.~Di Teodoro$^{15}$, B.-Q.~For$^{16,17}$, Tyler J.~Foster$^{18}$, J.~F.~G\'omez$^{6}$, H.~Imai$^{19,20}$, G.~Joncas$^{21}$, C.-G.~Kim$^{22}$, M.-Y.~Lee$^{23}$, C.~Lynn$^{1}$, D.~Leahy$^{24}$, Y.~K.~Ma$^{1,25}$, A.~Marchal$^{26}$, D.~McConnell$^{13}$, M.-A.~Miville-Desch\^{e}nes$^{9}$, V.~A.~Moss$^{13,27}$, C.~E.~Murray$^{15}$, D.~Nidever$^{28}$, J.~Peek$^{15,29}$, S.~Stanimirovi\'{c}$^{30}$, L.~Staveley-Smith$^{16,17}$, T.~Tepper-Garcia$^{17,27,31}$, C.~D.~Tremblay$^{4}$, L.~Uscanga$^{32}$, J.~Th.~van~Loon$^{33}$, E.~V\'{a}zquez-Semadeni$^{34}$, J.~R.~Allison$^{13,35}$, C.~S.~Anderson$^{36}$, Lewis Ball$^{8,13}$, M.~Bell$^{37}$, D.~C.-J.~Bock$^{13}$, J.~Bunton$^{13}$, F.~R.~Cooray$^{13}$, T.~Cornwell$^{13}$, B.~S.~Koribalski$^{13,38}$, N.~Gupta$^{13,39}$, D.~B.~Hayman$^{13}$, L.~Harvey-Smith$^{38,40}$, K.~Lee-Waddell$^{4,16}$, A.~Ng$^{13}$, C.~J.~Phillips$^{13}$, M.~Voronkov$^{13}$, T.~Westmeier$^{16}$, M.~T.~Whiting$^{13}$
\affil{$^1$Research School of Astronomy and Astrophysics, The Australian National University, Canberra, ACT 2611, Australia}
\affil{$^2$CSIRO Information Management and Technology, GPO Box 1700, Canberra, ACT 2601, Australia}
\affil{$^3$School of Natural Sciences, University of Tasmania, Hobart, TAS 7001, Australia}
\affil{$^4$CSIRO Space and Astronomy, PO Box 1130, Bentley, WA 6102, Australia}
\affil{$^5$Department of Astronomy, Yale University, New Haven, CT 06520, USA}
\affil{$^6$ Instituto de Astrof{\'i}sica de Andaluc{\'i}a, CSIC, Glorieta de la Astronom{\'i}a s/n, 18008 Granada, Spain}
\affil{$^7$Department of Physics, University of Sydney, Sydney, NSW 2006, Australia}
\affil{$^8$SKA Observatory, Jodrell Bank, Lower Withington, Macclesfield, Cheshire SK11 9FT, UK}
\affil{$^9$AIM, CEA, CNRS, Université Paris-Saclay, Université Paris Diderot, Sorbonne Paris Cité, F-91191 Gif-sur-Yvette, France}
\affil{$^{10}$Department of Physics, Stanford University, 382 Via Pueblo Mall, Stanford, CA 94305, USA}
\affil{$^{11}$Kavli Institute for Particle Astrophysics \& Cosmology, P.O. Box 2450, Stanford University, Stanford, CA 94305, USA}
\affil{$^{12}$Department of Physics and Astronomy and MQ Research Centre in Astronomy, Astrophysics, and Astrophotonics, Macquarie University, NSW 2109, Australia}
\affil{$^{13}$Australia Telescope National Facility, CSIRO Space and Astronomy, PO Box 76, Epping NSW 1710, Australia}
\affil{$^{14}$ASTRON - The Netherlands Institute for Radio Astronomy, 7991 PD Dwingeloo, The Netherlands}
\affil{$^{15}$Department of Physics \& Astronomy, Johns Hopkins University, 3400 N. Charles Street, Baltimore, MD 21218}
\affil{$^{16}$International Centre for Radio Astronomy Research (ICRAR), The University of Western Australia, 35 Stirling Hwy, Crawley, WA, 6009, Australia}
\affil{$^{17}$ARC Centre of Excellence for All Sky Astrophysics in 3 Dimensions (ASTRO 3D)}
\affil{$^{18}$Department of Physics \& Astronomy, Brandon University, 270-18 Street, Brandon, Manitoba, Canada}
\affil{$^{19}$Center for General Education, Comprehensive Institute of Education, 1-21-30 Korimoto, Kagoshima 890-0065, Japan}
\affil{$^{20}$Amanogawa Galaxy Astronomy Research Center, Graduate School of Science and Engineering, 1-21-30 Korimoto, Kagoshima 890-0065, Japan}
\affil{$^{21}$Universit\'{e} Laval, Pavillon Alexandre-Vachon 1045, Avenue de la Medecine, Canada}
\affil{$^{22}$Department of Astrophysical Sciences, Princeton University, Princeton, NJ 08544, USA}
\affil{$^{23}$Korea Astronomy and Space Science Institute, 776, Daedeokdae-ro, Yuseong-gu, Daejeon 34055, Republic of Korea}
\affil{$^{24}$Department of Physics and Astronomy, University of Calgary, Calgary, AB T2N 1N4, Canada}
\affil{$^{25}$Max-Planck-Institut f{\"u}r Radioastronomie, Auf dem H{\"u}gel 69, D-53121 Bonn, Germany}
\affil{$^{26}$Canadian Institute for Theoretical Astrophysics, University of Toronto, 60 St. George Street, Toronto, ON M5S 3H8, Canada}
\affil{$^{27}$Sydney Institute for Astronomy, School of Physics, The University of Sydney, NSW 2006, Australia}
\affil{$^{28}$Department of Physics, Montana State University, P.O. Box 173840, Bozeman, MT 59717-3840, USA}
\affil{$^{29}$Space Telescope Science Institute, 3700 San Martin Drive, Baltimore, MD 21218, USA}
\affil{$^{30}$Department of Astronomy, University of Wisconsin-Madison, 475 North Charter Street, Madison, WI 53706-15821, USA}
\affil{$^{31}$Centre for Integrated Sustainability Analysis, School of Physics, The University of Sydney, NSW 2006, Australia}
\affil{$^{32}$University of Guanajuato, A.P. 144, 36000, Guanajuato, Gto., Mexico}
\affil{$^{33}$Lennard-Jones Laboratories, Keele University, ST5 5BG, UK}
\affil{$^{34}$Instituto de Radioastronom\'{i}a y Astrof\'{i}sica, Universidad Nacional Aut\'{o}noma de M\'{e}xico, Apdo. Postal 3-72, Morelia, Michoac\'{a}n, 58089, M\'{e}xico.}
\affil{$^{35}$Sub-Dept. of Astrophysics, Department of Physics, University of Oxford, Denys Wilkinson Building, Keble Rd., Oxford, OX1 3RH, United Kingdom}
\affil{$^{36}$Jansky Fellow of the National Radio Astronomy Observatory, P. O. Box 0, Socorro, NM 87801, USA}
\affil{$^{37}$University of Technology Sydney, 15 Broadway, Ultimo, NSW 2007, Australia}
\affil{$^{38}$Western Sydney University, Locked Bag 1797, Penrith, NSW 2751, Australia}
\affil{$^{39}$Inter-University Centre for Astronomy and Astrophysics, Post Bag 4, Ganeshkhind, Pune 411 007, India}
\affil{$^{40}$Faculty of Science, UNSW Sydney, NSW 2052 Australia}
}

\begin{document}
\begin{frontmatter}
\maketitle
\end{frontmatter}
\begin{frontmatter}
\begin{abstract}
We present the most sensitive and detailed view of the neutral hydrogen ($\hi$) emission associated with the Small Magellanic Cloud (SMC), through the combination of data from the Australian Square Kilometre Array Pathfinder (ASKAP) and Parkes (Murriyang), as part of the Galactic Australian Square Kilometre Array Pathfinder (GASKAP) pilot survey. These GASKAP-HI pilot observations, for the first time, reveal $\hi$ in the SMC on similar physical scales as other important tracers of the interstellar medium, such as molecular gas and dust. The resultant image cube possesses an rms noise level of 1.1 K (1.6 mJy/beam) per 0.98 km s$^{-1}$ spectral channel with an angular resolution of 30$''$ ($\sim$10 pc). We discuss the calibration scheme and the custom imaging pipeline that utilizes a joint deconvolution approach, efficiently distributed across a computing cluster, to accurately recover the emission extending across the entire $\sim$25 deg$^2$ field-of-view. We provide an overview of the data products and characterize several aspects including the noise properties as a function of angular resolution and the represented spatial scales by deriving the global transfer function over the full spectral range. A preliminary spatial power spectrum analysis on individual spectral channels reveals that the power-law nature of the density distribution extends down to scales of 10 pc. We highlight the scientific potential of these data by comparing the properties of an outflowing high velocity cloud with previous ASKAP+Parkes $\hi$ test observations. 
\end{abstract}
\end{frontmatter}

\section{INTRODUCTION }
\label{sec:intro}

The evolution of the interstellar medium (ISM), which is driven by processes such as gas infall, star formation and subsequent stellar feedback, is fundamentally linked to the formation and overall evolution of galaxies. Stellar feedback in the form of outflows from sites of massive star formation greatly influences the overall anatomy of the ISM by means of the enrichment of heavy elements \citep{Nomoto2013}, mixing \citep{Krechel2020}, and transfer of kinetic energy \citep{Fierlinger2016}. In terms of gas infall --- which ultimately fuels star formation --- the accretion of diffuse gas onto the disks of galaxies is the leading explanation for how galaxies have continued to form stars while retaining a nearly constant atomic neutral hydrogen ($\hi$) content since $z\sim$2 \citep{noterdaeme2012, madau2014}. Hydrogen is characterized into several phases in models of the ISM (e.g., \citealt{wolfire95, wolfire03, vazquez12}), assuming solar metallicty: molecular (H$_2$; $T$ $\leq$ 50 K), cold neutral medium (CNM; $T$ $\sim$ 100 K), lukewarm neutral medium (LNM; 100 K$\leq T\leq$ 6000 K), warm neutral medium (WNM; $T\geq$ 6000 K), and ionized ($T\geq$ 8000 K). While hydrogen in all its phases dominates the composition of the ISM \citep{wolfire95, wolfire03, kalberla2018}, the densest and coolest regions that host and fuel site of star formation contain mostly molecular gas, including the CO-dark phase consisting of H$_2$ that lies outside carbon monoxide (CO) emitting regions (T$\sim$10 K; $n\sim$1000 cm$^{-3}$; \citealt{gnedin2009}).

There is an abundance of theoretical work that describes the astrophysics behind the life cycle of the multi-phase ISM (e.g., \citealt{McKeeOstriker1997, wolfire03, audit2005, vazquez12}); however, there are few observational studies that thoroughly explore the interconnected roles of dense molecular gas, dust, and the cold and warm neutral phases as traced by $\hi$ at equivalent spectral ($<$ 1 km s$^{-1}$) and physical pc scales. This paper introduces the first images from the $\hi$ pilot observations of the Galactic Australian Square Kilometre Array Pathfinder (GASKAP; \citealt{dickey2013}) Survey, which aims to explore the relationship between $\hi$ and other fundamental phases of the ISM within the Milky Way and nearby Magellanic System. 

The proximity of the Large and Small Magellanic Clouds (LMC and SMC), two nearby dwarf galaxies at the respective distances of 50 kpc and 60 kpc \citep{hilditch2005, pietrzynski2013}, makes them a unique laboratory for the detailed study of the impact from powerful outflows on the makeup of the surrounding ISM and thus the overall regulation of star formation. The GASKAP $\hi$ spectral line cubes possess the sensitivity, angular resolution, and spectral resolution to study the ISM at scales approaching the molecular gas observed with {\tt ALMA} ($\sim$0.5$''$ to 3$''$; 0.15 pc to 0.90 pc at the distance of the SMC) and comparable to dust as revealed by space-based telescopes in the far-IR such as the \textit{Spitzer Space Telescope} and the \textit{Herschel Space Observatory} ($\sim$20$''$ to 40$''$; 6 pc to 12 pc at the distance of the SMC). \citet{McClure-Griffiths2018} and \citet{dempsey2020} identified cool $\hi$ outflows around the SMC in emission and absorption, respectively. Furthermore, \citet{diTeodoro2019} detected for the first time $^{12}$CO gas entrained within one of these $\hi$ outflows at anomalous SMC velocities. The presence of molecular gas within outflowing $\hi$ at velocities exceeding the escape velocity suggests that the SMC could eject its entire cool gas reservoir that fuels star formation within $\sim$1--3 Gyr, while also enriching the surrounding circum-galactic medium and ultimately feeding this gas to the Milky Way. On the other hand, the reservoir of cold gas could be replenished by gas infall, which has been suggested for regions of the LMC from the Magellanic Bridge based on anomalous kinematic features in $\hi$ \citep{indu2015}. The pilot GASKAP observations presented here are ideal for the detailed study of both inflowing and outflowing $\hi$ and high velocity clouds (HVCs) around the SMC and the beginnings of the Stream and Bridge, which will ultimately inform important constraints on the future evolution of the Magellanic System and Milky Way. 
 
Turbulence is another key process that drives the evolution of the ISM \citep{elmegreen_scalo2004, MacLow2004} on a variety of length scales. For example, the accretion of circumgalactic material \citep{klessen_hennebelle2010}, galactic-scale gravitational instabilities, and magnetorotational instabilities likely inject turbulent energy on large scales ($>$ kpc;  \citealt{krumholz_burkhart2016, ibanesmej2017}), while stellar feedback from outflows and supernova explosions inject energy on smaller scales ($\sim$ 10 pc; \citealt{padoan2016, grisdale2017}). 
Similar to studies of star formation, there are a multitude of theoretical works that characterize turbulence over a range of length scales and density regimes \citep{kowal_lazarian2007, padoan2012, agertz2015}; however, few observational studies have attempted to probe turbulent properties in the context of a multi-phase ISM over similar physical scales \citep{stanimirovic1999, padoan2006, burkhart2010, pingel2013}. Specific to the SMC, \citet{szotkowski2019} used early pilot ASKAP observations to investigate spatial variations of the $\hi$ power spectrum and found the spectral slopes to be relatively uniform. \citet{kalberla19} investigated the turbulent properties of $\hi$ separated into the CNM, LNM, and WNM to reveal that each phase has distinct turbulent properties. \citet{eden21} analyzed the two-dimensional spatial power-spectrum of maps of the dense gas mass fraction derived independently from a combination of H$_2$ column density estimated from dust continuum and the ratio of CO intensities to show $\sim$10 pc is the characteristic scale for the largest variations in the clump and star formation efficiencies in CO-traced clouds formed by supersonic turbulence. \citet{pingel18a} explicitly compared the turbulent properties of several multi-wavelength tracers in the Perseus molecular cloud including dust, $\hi$, and CO. Each tracer showed characteristics of a distinct turbulent environment, such as self-gravitating, supersonic medium in the dust and mostly transonic medium in the $\hi$. Each of these studies were able to draw conclusions on how turbulence influences the structure and dynamics of different ISM phases. However, a true multi-phase characterization of ISM turbulence has been limited by differences in the angular and spectral resolutions of observational data.


The GASKAP survey is designed as a high angular and spectral resolution ASKAP survey of the Milky Way Galactic plane and the Magellanic System in $\hi$ and hydroxyl (OH) spectral lines at frequencies of $\lambda = 21~{\rm cm}$ and $18~{\rm cm}$. The $\hi$ component of the GASKAP survey (hereafter referred to as GASKAP-HI) will probe the properties of $\hi$ in our own Galaxy and the nearby Magellanic System at unprecedented spatial resolution ($\sim$10 pc at the distance of the SMC), similar to the scale of individual dark and star forming regions in Galactic molecular clouds (e.g., Perseus, \citealt{young-lee_2012}) and spectral ($\sim$0.5 km s$^{-1}$) resolutions. The OH component of GASKAP will serve as an important probe of the molecular content of the ISM not associated with CO emission \citep{Allen2012, Allen2015, Busch2021}. The survey's primary objective is to follow the cycle of gas evolution from diffuse, warm \hi\ through cold \hi, to molecular gas as traced by OH, and finally to star formation and evolution through OH masers \citep{dickey2013}.



Many previous surveys have looked at the $\hi$ in the Galaxy and Magellanic System. However, previous large-area single dish surveys of the $\hi$ \citep{McClure-Griffiths2009, kerp2011, peek2018, Martin2015, HI4PI} have all lacked sufficient spatial resolution to probe the structure and dynamics at small scales. Interferometers, on the other hand, sacrifice surface brightness sensitivity \citep{braun1985, stanimirovic2002} for high angular resolution in order to reveal bright, small-scale structures with narrow velocity widths. GASKAP-HI builds on the Canadian Galactic Plane Survey \citep{taylor2003}, Southern Galactic Plane Survey (SGPS; \citealt{McClure-Griffiths2005}), VLA Galactic Plane Survey (VGPS; \citealt{Stil2006}), the HI/OH Recombination-line Survey of the Inner Milky Way (THOR; \citealt{beuther2016}), the Southern Galactic Centre Survey \citep{McClure-Griffiths2012}, and surveys of the Magellanic System \citep{kim1998, staveley-smith1997, stanimirovic1999} by increasing the angular resolution to arcsecond scales (a factor of two increase relative to SGPS and VGPS), increasing the spectral resolution to 0.5 km s$^{-1}$ (useful to probe kinetic temperatures of $\sim$6 K), and expanding the instantaneous $uv$-coverage to characterize a virtually continuous range of spatial frequencies. Combining these novel GASKAP-HI data with data from the 64 m Parkes single dish telescope (Murriyang) will allow, for the first time, a view of nearby $\hi$ emission at similar physical scales as dust and molecular gas in the Magellanic System and supplement the aforementioned surveys of the Galactic Plane. Furthermore, coupling the detailed view of the $\hi$ emission structure with high spectral resolution $\hi$ absorption measurements will facilitate accurate measurements of the spin temperature, placing lower limits on the kinetic temperatures of the WNM and LNM. The high number densities in CNM thermalise the $\hi$ line in this phase such that absorption measurements will provide the true kinetic temperature.

Each antenna element of the Australian Square Kilometer Array Pathfinder (ASKAP; \citealt{hotan2021}) telescope is outfitted with a Phased Array Feed (PAF) receiver and a matched set of digital equipment that forms 36 simultaneous primary beams on the sky, expanding the instantaneous field of view (FoV) from $\sim$1 deg$^2$ to $\sim$25 deg$^2$. The unique wide-field imaging capability of ASKAP presents several unique data processing challenges that the GASKAP-HI survey must address before beginning full survey operation. For example, it is paramount to remove the inherent instrumental point-spread function (PSF) from a sky brightness distribution that extends over the boundaries of multiple beams without degrading the recovered structure. It is also crucial to combine ASKAP data with data from a single dish to fill in the missing short-spacings filtered out by the fixed antenna distribution \citep{stanimirovic1999, stanimirovic2002}.


ASKAP surveys, including GASKAP, were selected in 2009 based on scientific merit during the construction phase of the telescope. The plans and specifications of each survey have been progressively refined in the intervening years to match the telescope capabilities and increased computing power, as described in \citet{hotan2021}. In late 2019, ASKAP commenced pilot observations for each survey, allocating 100 hours of observing time to each survey team to test full survey operations. In this paper, we present the first GASKAP-HI pilot observations of the SMC, obtained in December 2019. The observations are one part of a multi-faceted pilot survey, covering an extended region of the SMC, the starts of the Magellanic Stream and Bridge, as well as two fields in the Milky Way Galactic Plane. These images are also the first to be produced by our custom imaging pipeline built around {\tt WSClean}, a command-line application first introduced by \citet{offringa2014} that is specifically designed to perform efficient wide-field imaging. This software allows us to perform joint deconvolution distributed over a computing cluster, which ensures accurate reconstruction of the extended $\hi$ emission intrinsic to the Galaxy and Magellanic System while also managing the overall memory footprint.

This paper is structured as follows: Section~\ref{sec:obs} describes the philosophy and parameters of the GASKAP-HI pilot survey; Section~\ref{sec:data_reduction} explains the calibration procedure, provides a thorough description of our custom imaging pipeline, including the combination with Parkes data to correct for the missing short-spacings, and discusses our quality assessment methods; Section~\ref{sec:data_products} outlines the noise properties of our final image cubes and derives the global transfer function to fully characterize the represented spatial frequencies; Section~\ref{sec:hvc} highlights the advantages of the increased sensitivity and angular/spectral resolution through global channel maps of the SMC and presents an analysis of a known outflowing high velocity cloud; finally, in Section~\ref{sec:summary}, we summarize these results from the SMC and anticipate future results from our other pilot fields and the eventual full GASKAP-HI survey.

\section{ASKAP Observations and the GASKAP-HI Pilot Survey}\label{sec:obs}
GASKAP-HI makes optimum use of ASKAP's wide field-of-view and bi-modal antenna baseline distribution with peaks near 500 m and 3000 m to simultaneously image the diffuse emission with high surface brightness sensitivity and high angular resolution for point sources, such as $\hi$ absorption against continuum sources. GASKAP-HI will give a detailed view of gas evolution in galaxies by targeting the three very different galaxies: the Milky Way, the LMC and SMC.


GASKAP is unique amongst the ASKAP surveys in its requirement for high spectral resolution. The standard ASKAP spectral resolution is $\Delta \nu= 1~{\rm MHz}$ for the continuum surveys (EMU; \citealt{norris2011}; POSSUM \citealt{murphy2013}; VAST; \citealt{anderson2021}) and $\Delta \nu =18.5~{\rm kHz}$, 
optimised for extragalactic \hi\ surveys like WALLABY \citep{korbibalski2020}, DINGO \citep{meyer2017}, and FLASH \citep{allison2020}. By contrast, GASKAP is designed to fully resolve the spectral linewidths observed in cold $\hi$, OH masers, and OH absorption, requiring frequency resolution of  $\Delta\nu \lesssim 5~{\rm kHz}$. This is achieved through a `zoom' mode of the ASKAP poly-phase filterbank, through which observations can be obtained with one of several different spectral resolutions between $\Delta \nu = 0.58~{\rm kHz}$ and the standard resolution of $\Delta \nu =18.5~{\rm kHz}$.  GASKAP-HI plans to use a spectral resolution of $\Delta \nu = 1.15~{\rm kHz}$, giving a velocity resolution of $\Delta v \approx 0.3~{\rm km~s^{-1}}$ over a total bandwidth of $18~{\rm MHz}$ spanning 15552 fine channels. We inclusively select the fine channels 7887 to 9934 $\sim$400 km s$^{-1}$ to $-100$ km s$^{-1}$) in the topocentric reference frame, which captures $\hi$ emission from the Magellanic System and Milky Way, to reduce the total amount of data that needs processing. The center frequency of each measurement set is 1419.81 MHz in the topocentric reference frame. 


Our observations of the SMC total 20.9 h of integration, split equally into two separate sessions. Each session is known as a schedule block (SB) and identified with IDs (SBIDs) 10941 and 10944, making it easier to identify observations in the CSIRO ASKAP Science Data Archive (CASDA\footnote{\url{https://data.csiro.au/collections/domain/casdaObservation/search/}}). The PAF footprint, known as closepack36, was centered on J2000 RA = 00$^{\rm h}$58$^{\rm m}$43.280$^{\rm s}$, Dec = $-$72$^{\rm d}$31$^{\rm m}$49.03$^{\rm s}$. This particular PAF footprint places 36 simultaneously formed beams in a 6$\times$6 hexagonal grid such that the beam centers are separated by 0.9 deg to provide more uniform sensitivity \citep{McConnell2016, hotan2021}. Throughout the observation, every 10 minutes the PAF footprint cycles through three pointing centers provided in Table~\ref{tab:obs_params}. This process, known as interleaving, ensures uniform sensitivity across the FoV. Due to beams formed offset from the pointing center, it is necessary to track the parallactic angle on the sky throughout the interleaving process. The ASKAP antennas achieve this through a unique third axis of rotation on the reflector itself, in addition to a simple azimuth-elevation mount. Table~\ref{tab:obs_params} summarizes the observational setup. 

\begin{table*}
\centering
\caption{Summary of GASKAP-HI pilot observations. $t_{\rm int}$ is the integration time spend on each interleave position (20.9 h total integration), $\nu_0$ is the central frequency of our measurement sets (see text) in a topocentric frame, $BW$ is the total bandwidth, $B$ is the native spectral resolution, and $\theta_{\rm PAF}$ is the PAF rotation, which refers to the total rotation of the footprint on the sky, including the natural $-$45 deg rotation to align with celestial coordinates. Note that the pointing centers were cycled every 10 minutes throughout the listed UTC date and time ranges.}
\label{tab:obs_params}
\resizebox{\textwidth}{!}{
\begin{tabular}{lccccccccc}
\hline
Interleave & $t_{\rm int}$ [h] & $\nu_0$ [MHz] & $BW$ [MHz] & $B$ [kHz] & J2000 & Galactic & UTC Date \& Time Range (SB10941) & UTC Date \& Time Range (SB10944) & $\theta_{\rm PAF}$ [deg]  \\
\hline
A &  6.97 & 1419.81 & 18.5 & 1.157 & RA = 00$^{\rm h}$58$^{\rm m}$43.3$^{\rm s}$, Dec = $-$72$^{\rm d}$31$^{\rm m}$49.0$^{\rm s}$ &  $l$=302.18$^{\circ}$, $b$=$-$44.59$^{\circ}$ & 2019 Dec 22 06:12:04.8 to 16:07:37.1 & 2019 Dec 23 06:06:40.8 to 16:16:39.0 & $-$48.0 \\
B &  6.97 & 1419.81 & 18.5 & 1.157  & RA = 01$^{\rm h}$04$^{\rm m}$26.2$^{\rm s}$, Dec = $-$72$^{\rm d}$14$^{\rm m}$31.6$^{\rm s}$ &  $l$=301.53$^{\circ}$, $b$=$-$44.85$^{\circ}$ & 2019 Dec 22 06:22:21.9 to 16:17:54.2 & 2019 Dec 23 06:16:57.9 to 16:31:44.7 & $-$49.4 \\
C &  6.97  & 1419.81 & 18.5 & 1.157 & RA = 01$^{\rm h}$04$^{\rm m}$58.2$^{\rm s}$, Dec = $-$72$^{\rm d}$45$^{\rm m}$36.6$^{\rm s}$ &  $l$=301.53$^{\circ}$, $b$=$-$44.33$^{\circ}$ & 2019 Dec 22 06:32:39.0 to 16:28:01.3 & 2019 Dec 23 06:27:05.0 to 16:41:51.9 & $-$49.5  \\

\hline
\end{tabular}}
\end{table*}

\section{Data Reduction}\label{sec:data_reduction}

\subsection{ASKAPsoft Pipeline}\label{subsec:askapsoft}

ASKAP observations are calibrated and imaged using a high-performance processing pipeline \citep{2020ASPC..522..469W,hotan2021} that runs on the \textit{Galaxy} supercomputer at the Pawsey Supercomputing Centre. This typically includes all of the steps necessary to calibrate the data and produce images and spectral cubes of the observation, using the \texttt{ASKAPsoft} package \citep{2019ascl.soft12003G}. For the GASKAP-HI processing, since the imaging is done in the joint-imaging pipeline described below, the ASKAP pipeline was limited to bandpass calibration, flagging, and self-calibration of the time-dependent gains.

\subsubsection{Bandpass and Flux Scale Calibration and Flagging}\label{subsubsec:bandpasscal}

Prior to self-calibration, the ASKAP pipeline prepares the data through calibration of the bandpass and flagging of unwanted signal. The bandpass and overall flux scale is calibrated using the primary flux calibrator PKS B1934$-$638. The target and flux calibrator data are then flagged to remove radio-frequency interference (RFI). This employs a dynamic flagging algorithm operating in the time-domain for each channel, which flags discrepant time ranges based on the Stokes V flux.

Data is then averaged to form 1~MHz-wide channels (that is, averaging 864 channels together), which will be used to create the continuum images used in the self-calibration. Those channels that are identified to contain HI emission are then flagged for this step, so that only continuum emission contributes to the resulting images.

\subsubsection{Self-calibration}\label{subsubsec:selfcal}
The derived beamforming weights are generally updated once every day or two and require 2.5 hours of telescope overhead for calibration \citep{hotan2021}. In lieu of spending time on traditional phase calibration methods --- such as regularly observing a reference source near the science target --- the amount of flux within the wide ASKAP FoV at any given time ensures adequate flux to utilize an in-field self-calibration scheme to correct for time-varying phase errors. The {\tt ASKAPsoft} pipeline creates continuum images for each beam from the calibrated, flagged, and averaged science visibilities to use as initial models. No modeling of the spectral index is performed due to the relative narrow bandwidth.    

Due to the relatively weak continuum in this field, no baselines were excluded in the continuum imaging. For future GASKAP-HI fields on the Galactic Plane or LMC, baselines below 500 m will be excluded. An image of the field is produced with the {\tt ASKAPsoft} BasisfunctionMSF solver with Hogbom clean and a shallow sky model of the field is made from that image. A total of 257 w-planes are used in this image. This model is used to calibrate the complex gains over short intervals ($\approx 30$ sec) across the entire observation. To preserve the flux scale, we perform phase-only self-calibration by fixing the amplitude of the calibration parameters to 1. The resulting calibration parameters are applied to both the continuum and spectral visibility data sets, and a subsequent continuum image is made using the calibrated data. 

\subsection{Observation Diagnostics}
The size of the calibrated visibilities from a typical GASKAP-HI field, consisting of roughly 20 hours of integration time, 2048 spectral channels, and 108 total beams (36 beams$\times$3 interleaves) in closepack36 configuration, totals $\sim$ 7 TB. Before expending the large effort required for spectral line imaging to create an 11 GB image cube, it is important to understand the quality of an observation.
This allows us to verify that the observation is suitable for its intended science use.
In particular, we need to know whether
\begin{itemize}
    \item the sensitivity across the field is even, so that features can be compared across the field;
    \item there are enough data available from short baselines to provide good sensitivity to large scale structure and low surface brightness features;
    \item there are enough data available from long baselines to adequately sample fine structure and to detect absorption;
    \item the calibration bandpasses are reasonably free of structure.
\end{itemize}
We produce a diagnostics report to describe the quality of the observation.\footnote{\url{https://casda.csiro.au/validation/10941/AS108/validation_gaskap_10941/} and \url{https://casda.csiro.au/validation/10944/AS108/validation_gaskap_10944/}}
It summarises and visualises outputs produced by the {\tt ASKAPsoft} \citep{2019ascl.soft12003G} pipeline up to the production of calibrated measurement sets, along with the continuum image and source catalogue of the field. The report provides a snapshot of the overall quality and is generally sufficient to determine whether to proceed with imaging or re-observe.

The first two sections of the report provide details of the observation and the continuum image. 
In the observation we need to know where the field is and how it was observed --- duration, beam pattern (or PAF footprint), and the correlator mode.
The footprint is also used later to determine the layout of beam based plots.
Noting that the continuum data is well covered by the ASKAP continuum validation report, the continuum image report is limited to basic details.

To allow assessment of the calibration we visualise the median calibration bandpasses both by beam and by antenna.
In the plots we highlight up to 10 bandpasses that either have
\begin{enumerate}[label=\alph*)]
\item a median more than one standard deviation from the median of all bandpasses, or
\item an amplitude range that is more than one standard deviation from the median amplitude range of all bandpasses.
\end{enumerate}
The first rule highlights antenna sensitivity issues and the second calls out structure in the calibration bandpasses.
As the raw ASKAP visibilities are not retained to manage disk space, recalibration is not possible, so any problems reflected here would generally mean the dataset needs to be rejected and the field re-observed.

\begin{figure*}
	\includegraphics[width=\textwidth]{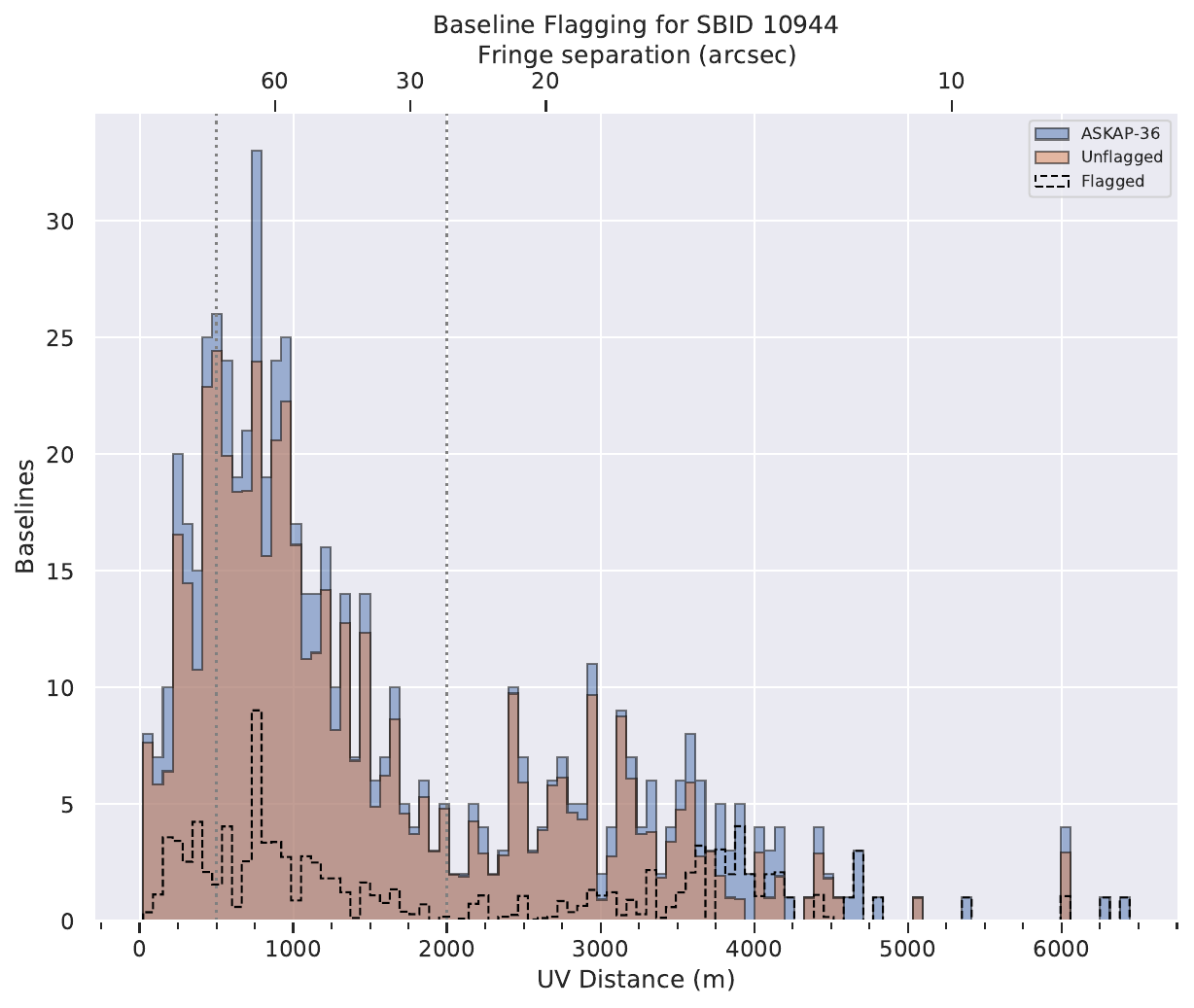}
    \caption{Baseline coverage for the ASKAP-36 array showing the proportion of baselines flagged and unflagged in the observation. The flagging proportion is measured across all beams and all channels. However, any time periods where the data is entirely flagged are excluded. The blue profile shows all available baselines, while the transparent orange profile denotes unflagged baselines. A histogram of the flagged baselines (excluding the automatically flagged autocorrelations) is outlined by the dotted line. The two vertical dotted lines represent the definition for short ($l_{\text{baseline}} < 500$ m) and long baselines ($l_{\text{baseline}} > 2000$ m)}
    \label{fig:baselines}
\end{figure*}

In the diagnostics section, we largely focus on what data have been flagged by the ASKAPSoft pipeline (see Section \ref{subsubsec:bandpasscal}) within the measurement sets and the effect this has on the suitability of the dataset for different science cases.
The flagging of the measurement sets is done in four dimensions: by baseline (or combinations of antennas), by beam, by channel, and by time.
Flagging is generally used to remove bad data from the dataset, whether for an antenna that is performing incorrectly, an antenna that is in the shadow of another antenna with respect to the target source, or for radio frequency interference.
We provide plots of the flagging 
\begin{enumerate}[label={\alph*)}]
\item by baseline, (see Figure \ref{fig:baselines}), which shows if there is any bias in coverage by angular scale, 
\item by time to show if our u-v coverage is even, and 
\item by percentage of each baseline flagged to highlight frequency based flagging.
\end{enumerate}For example, the histogram of the baseline coverage across all beams and all channels for the entire observation for SBID 10944 in Figure~\ref{fig:baselines} shows a majority of the available baselines remain unflagged during this scheduling block. Figure~\ref{fig:baselines} also demonstrates that flagged baselines are evenly distributed across the entire range of available baselines. This ensures there are no unexpected gaps in the continuous $uv$-coverage, which would lead to potential artifacts at particular spatial scales in the final images. The two distinct peaks between 400 and 1200 m, and 2 km and 3 km are ideal for synthesised beam sizes between 30$''$ to 60$''$  to detect Galactic and Magellanic $\hi$ in emission. The longest baseline of 6 km produces a beam size of $\sim$10$''$ that is necessary for absorption measurements. All diagnostic plots are sourced from the flagging summary files for each measurement set.
As is done in the WALLABY diagnostics \citep{for2019, korbibalski2020}, the linked diagnostic reports also provide plots of 
\begin{enumerate}[label={\alph*)}]
\item the fraction flagged by beam, 
\item the number of antennas flagged in each beam and from these, 
\item the expected RMS per channel for each beam of the observation.
\end{enumerate}

To assist the assessment of quality of the observation we report five metrics on a good, uncertain, and bad scale.
\begin{enumerate}
\item The percentage of data flagged from short baselines ($l_{\text{baseline}} < 500$ m).
These are crucial for large scale emission and low surface brightness sensitivity.
\item The percentage of data flagged from long baselines ($l_{\text{baseline}} > 2000$ m).
These baselines are essential for tracing fine detail and for detecting absorption.
\item The percentage difference of the expected RMS across the field:
\begin{equation}
 \text{difference} = \frac{\rm RMS_\text{exp,max}-RMS_\text{exp,min}}{\rm RMS_\text{exp,min}}
\end{equation}
which is used to show how even the sensitivity is across the field.
\item The percentage of integrations with less than 5\% of channels flagged. In observations of local \hi\ we do not want the diffuse emission spectral lines flagged out.
\item The number of unflagged antennas with a phase RMS $\geq 40^\circ$. 
These are antennas where the phase solutions are severely unstable over the course of the observation, leading to poor positional accuracy.
\end{enumerate}
These metrics are made available both in the report and on CASDA \citep{2020ASPC..522..263H}. The metrics and their threshold values are shown in Table \ref{tab:diagnostics}.
For an observation to be acceptable for emission studies, both the short baseline and the expected RMS variance tests must be in the uncertain or good ranges.
For an observation to be acceptable for absorption studies, both the long baseline and the expected RMS variance tests must be in the uncertain or good ranges.

\begin{table*}
\centering
\caption{Observation Diagnostics Metrics}
\label{tab:diagnostics}
\begin{tabular}{lcccl}
\hline
Metric & Good & Uncertain & Bad & Units \\
\hline
Flagged Short Baselines & $< 20$ & $20 - 40$ & $\geq 40$ & percent \\
Flagged Long Baselines & $< 30$ & $30 - 45$ & $\geq 45$ & percent \\
Expected RMS Variance & $< 10$ & $10 - 30$ & $\geq 30$ & percent \\
Unflagged Integrations & $> 70$ & $70 - 50$ & $< 50$ & percent \\
Bad Phase Antenna & $0$ & $1 - 2$ & $\geq 3$ & antenna \\
\hline
\end{tabular}
\end{table*}

Based on these metrics, we find that the observations presented here are suitable to use for emission line analysis. Whilst antennas (denoted by the prefix `ak') ak06 and ak32 are completely flagged out, the flagging is even across baseline lengths, with flagging percentages of ~16\% on short baselines and ~25\% on long baselines (denoted by the dashed lines in Figure~\ref{fig:baselines}). There is increased flagging on integrations around the change of interleaves. 
In scheduling block 10941, the last integrations for the observation in all interleaves are heavily flagged. The calibration bandpasses exhibit structure with small amplitudes, including dips at the 1 MHz beam forming intervals. These dips are not an issue for GASKAP-HI due to our relatively narrow bandpass.

\subsection{Joint-Imaging Pipeline}\label{subsec:casapipeline}
The PAF receivers equipped on each ASKAP antenna extend the typical FoV for a 12 m diameter antenna from $\sim$1 deg$^2$ to $\sim$25 deg$^2$ by simultaneously forming 36 distinct primary beams. Standard full-field ASKAP images are produced by deconvolving the inherent PSF response from the sky-brightness distribution of each separate beam independently and then linearly mosaicking them together. This approach works for most other ASKAP surveys, as the emission is small scale (e.g., distant galaxies with small angular size) and generally is contained within a single beam. Such techniques work less well for GASKAP-HI because the targeted diffuse $\hi$ emission extends up to the boundaries of the primary beam. To ensure we accurately recover the distribution of structure down to baselines corresponding to the size of the primary reflectors themselves (i.e., 12 m), fill in gaps between all the short baselines, and deconvolve the images deeper, we must utilize a joint deconvolution approach. Here, a large image is created from the ourier inversion of the sampled visibilities from all beams (i.e., the `dirty image') before proceeding with deconvolution. Since the calibrated visibilities from all of the beams must be managed simultaneously in a joint deconvolution approach, the computation requirements --- specifically the memory footprint --- are immense relative to other imaging techniques. As of publication, {\tt ASKAPsoft 1.3.0}\footnote{\url{https://www.atnf.csiro.au/computing/software/askapsoft/sdp/docs/current/general/releaseNotes.html}}, the primary processing software utilized to image all other ASKAP data, is not fully optimised to run a distributed joint deconvolution scheme. 

\begin{figure*}
\centering{
\includegraphics[width=3.25 in]{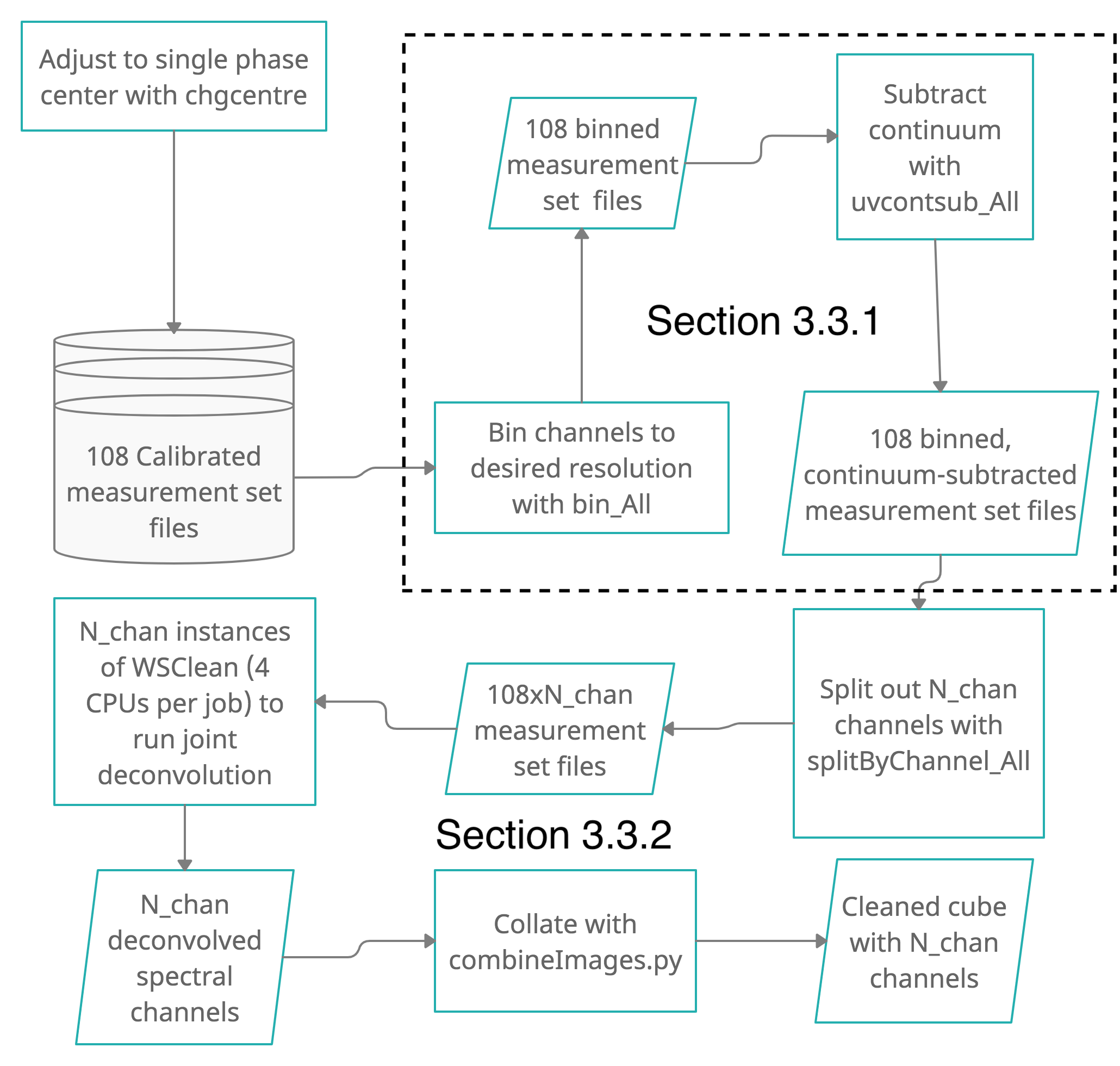}
\caption{\label{fig:pipelineFlowChart} A flow chart summarizing the workflow of our custom pipeline used to image a typical GASKAP-HI field. In brief, the phase centers for each of the 108 measurement sets, which contain the gain-calibrated visibilities per beam (36 beams$\times$3 interleaves), are set to single value using the {\tt chgcentre} utility available through the {\tt WSClean} package. These altered measurement sets can be binned and put through a $uv$-based continuum subtraction scheme, respectively using the CASA tasks {\tt split} and {\tt uvcontsub} within the scripts {\tt bin\_All} and {\tt uvcontsub\_All}, that is submitted in a distributed manner for efficient processing. Single spectral channels are then split out to ensure reasonable memory management. A series of batch jobs, each utilizing 4 total CPU cores, are submitted with the 108 measurement sets per spectral channel as input to the {\tt WSClean} command-line imager to produce a jointly deconvolved image. The final deconvolved images are then collated into a single cube before combination with Parkes data to fill in the missing short-spacings (see Section~\ref{subsubsec:combination}). The pipeline processes are represented as rectangles, actual input/output of these processes are represented by parallelograms, and the action of input is represented by the curved arrows.}}
\end{figure*}

We instead developed a custom imaging pipeline\footnote{\url{https://github.com/nipingel/GASKAP_Imaging}} that utilizes a combination of Common Astronomy Software Application ({\tt CASA}; \citealt{mcMullin2007}) tasks to pre-process the measurement sets, {\tt WSClean}\footnote{\url{https://gitlab.com/aroffringa/wsclean}} to grid the multiple ASKAP pointings onto a single grid for joint deconvolution using corresponding primary beam models for each ASKAP primary beam, and {\tt miriad} \citep{sault95} tasks to combine ASKAP and Parkes data to correct for short-spacings. This imaging package enables efficient gridding of visibilities and stable deconvolution, as demonstrated by the high-fidelity images produced of the Galactic Centre with the MeerKAT telescope \citep{heywood2019}. We input the ASKAP primary beams measured from holography at 1.4 GHz \citep{hotan2021} to ensure highly accurate gridding of the gain-calibrated visibilities and scaling of the components subtracted from the dirty image during deconvolution major and minor iterations, respectively. The pipeline takes advantage of a high-performance computing (HPC) environment, while also managing the memory footprint, by submitting simultaneous batch imaging jobs of individual channels. A majority of the processing was performed with the Avatar computing cluster located at the Australian National University Research School of Astronomy and Astrophysics. The configuration of {\tt WSClean} imaging mode and resource allocation is performed by bash scripts specific to both the Portable Batch System (PBS) and Simple Linux Utility for Resource Management (SLURM) scheduling software. 

Figure~\ref{fig:pipelineFlowChart} summarizes the work flow of our joint-imaging pipeline to image a single GASKAP-HI field. We download from CASDA 108 measurement set files that contain the gain-calibrated visibilities for a field observed in the closepack36 beam configuration. We then use the built-in {\tt WSClean} tool {\tt chgcentre} to phase-rotate the visibilities to Pointing center 1 listed in table~\ref{tab:obs_params}, thus setting a common direction where all phases are zero and effectively sets the center of the resultant image. To avoid smearing and noncoplanar baseline effects (see \citealt{cornwell2005}), such as the distortion of sources sitting far away from the phase center, the phase information in the measurement sets is updated during gridding based on the individual beam phase centers provided to {\tt WSClean} via a configuration file. 


\subsubsection{Binning and uv-based Continuum
Subtraction}\label{subsubsec:binningandcontsub}

\begin{figure}
\centering{
\includegraphics[width=3.25 in]{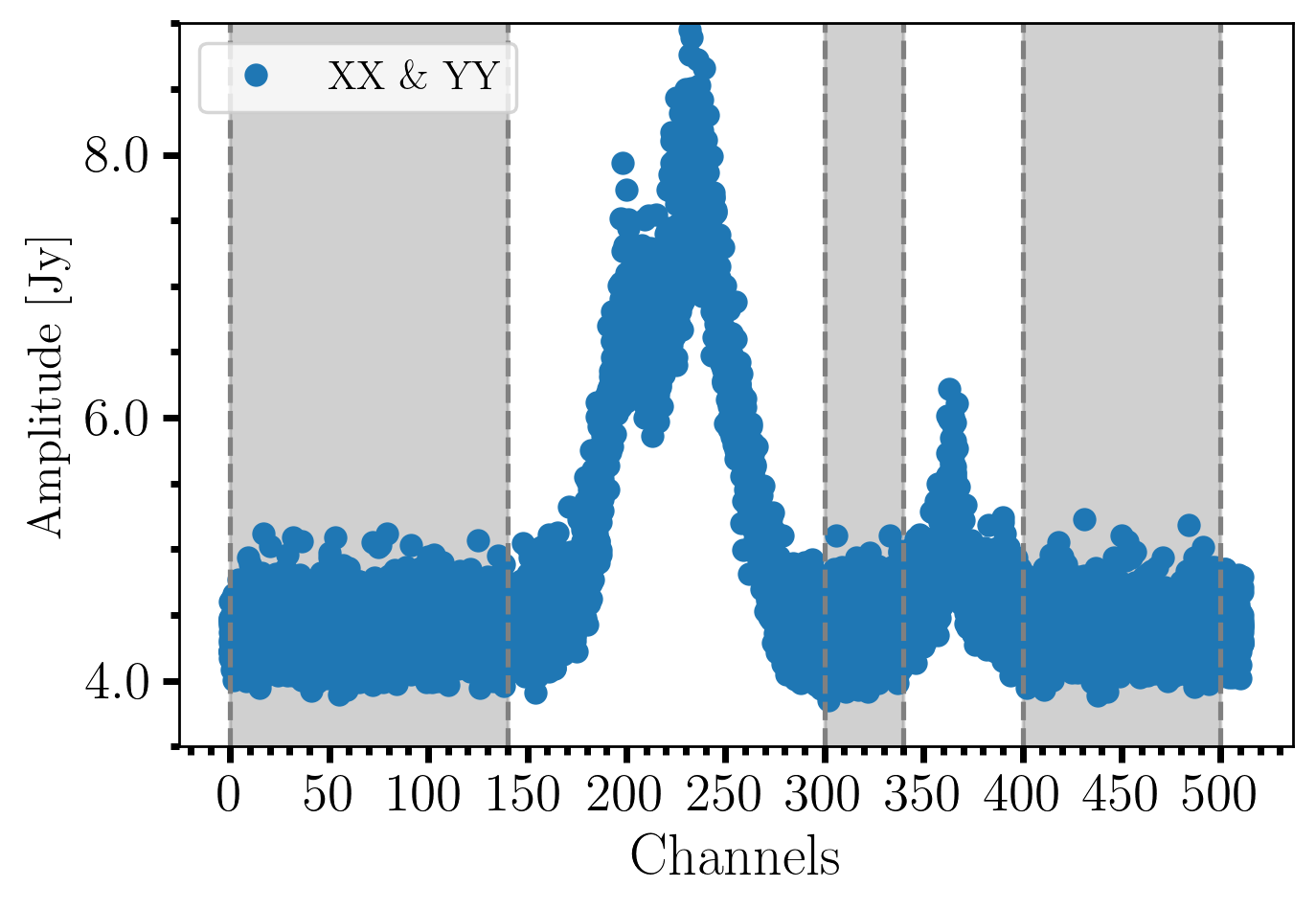}
\caption{\label{fig:example_spectrum_contsub} The mean amplitude of XX and YY correlations averaged over 10000 s as a function of binned channel for a central ASKAP beam taken from the baseline between ak04 and07. The $\hi$ emission-free channel range used to fit and subtract a first order polynomial model of the continuum is denoted by the shaded regions.}}
\end{figure}

For the purpose of increasing the signal-to-noise ratio (SNR), we first bin each measurement set by four channels, achieving a spectral resolution of 4.628 kHz ($\sim$0.98 km s$^{-1}$ at the rest frequency of $\hi$) for a single channel.

It is also necessary to remove the contribution from continuum sources in the field to ensure we infer the correct physical properties of the $\hi$ gas and to aid deconvolution. We select a range of $\hi$ emission-free channels by plotting a mean spectrum from the visibilities contained in several central beams and baselines between core antennas, which is expected to contain the brightest emission. An example spectrum is shown in Figure~\ref{fig:example_spectrum_contsub}. The continuum is removed from the $uv$ spectra for each individual beam by fitting and subtracting a first order polynomial to both the real and imaginary visibility spectra over the binned channel range from 0 to 140 (1418.63 MHz to 1419.28 MHz), 300 to 340 (1420.01 MHz to 1420.20 MHz), and 400 to 500 (1420.48 MHz to 1420.94 MHz), as no emission from the SMC or foreground Milky Way is observed in these channels. 


\subsubsection{Joint Deconvolution}\label{subsubsec:jointimaging}

\begin{table}
\centering
\caption{WSClean parameters. See \url{https://wsclean.readthedocs.io/_/downloads/en/latest/pdf/} for comprehensive documentation on these parameters.}
\label{tab:tclean_params}
\begin{tabular}{lc}
\hline
Parameter & Value [units]  \\
\hline
size & 5000x5000 [pixels] \\
scale & 7 [$''$] \\
niter & 25,000 \\
mgain & 0.7 \\
multiscale-scales & 0, 8, 16, 32, 64, 128, 256 \\
multiscale-scale-bias & 0.85 \\
weight & Briggs \\
robust parameter & 0.9 \\
threshold & 5 [mJy] \\
beamsize & 30 [$''$]  \\ 
\hline
\end{tabular}
\end{table}

Individual channels are split out from each of the 108 measurement set files to best manage the large memory footprint required by joint deconvolution. The imaging of each channel is submitted as a separate batch job that utilizes a total of 4 cores to take advantage of the built-in parallelization of multiple compute threads. Table~\ref{tab:tclean_params} summarizes the typical configuration to process a single channel with significant emission. We perform a multiscale CLEAN deconvolution \citep{cornwell2008} with: a pixel size (i.e., scale parameter) of 7$''$, multi-scale scales set to 0, 8, 16, 32, 64, 128, and 256 pixels (angular range between 0$''$ to 30$'$), the total number of iterations (niter) set to 25,000, and mgain set to 0.7, meaning that the peak flux must be reduced by 70\% to trigger the next major iteration; typical runs go through two to three major iterations. 

We set the auto-threshold parameter to 3, which sets the global stopping threshold as 3 times the measured background noise. This value is automatically computed at the end of each major iteration by measuring the median absolute deviation (MAD) of the noise. In general, these parameters allow the deconvolution to run until the peak residual reaches the global stopping threshold. We set this threshold to 5 mJy, which is $\sim$2.5 times the expected noise level, to avoid cleaning too deeply into the noise. The `smallscalebias' parameter is set at 0.85 to bias the multiscale deconvolution towards the smaller scales and aid cleaning of the expected small-scale features, and the gridded visibilities are weighted (without excluding any baselines) with a Briggs weighting scheme with a robust parameter of 0.9 to generate a PSF with higher sensitivity but still reasonable sidelobe levels. 

The final deconvolved images are spatially smoothed such that the final restoring beam is 30$''\times$30$''$ as a compromise between sensitivity to extended structure and final angular resolution. This is comparable to the synthesized beam size expected from our robust weighting applied to the $uv$ coverage. The total time to image a typical channel ranged between 4 to 8 hours, depending on the complexity of the emission present in the channel and number of major iterations. We find that the processing is generally memory limited, though processing of other pilot fields will increase optimisation and determine clear limitations.




\subsubsection{Filling in the Missing Short-Spacings}\label{subsubsec:combination}
The size of the primary reflectors on each antenna is an inherent physical limitation of how closely individual elements of an interferometer can be placed, thus filtering out the low spatial frequencies corresponding to the dish diameter (and below) and eliminating sensitivity to large angular scales. These missing baselines are commonly referred to as the missing short-spacings. Observations of the same area of the sky with a large single dish telescope with sufficient overlap in the $uv$-plane can be used to fill in the baseline samples below that of the minimum spacing between antennas \citep{stanimirovic2002} (i.e., the shortest baseline), including the total power at $u$ = $v$ = 0. The total flux and structure of diffuse emission on scales larger than that sampled by the shortest baseline is then recovered in the final map.

Several established techniques are available to combine interferometric and single dish observations either in the $uv$-plane after deconvolution (e.g., \citealt{stanimirovic2002, cotton2017}), the image plane directly \citep{faridani2018} or through approximating the single dish data as artificial visibilities to be included in the image reconstruction process applied to interferometer data \citep{koda2011, koda2019}. More recently, \citet{rau2019} have developed a generic joint reconstruction algorithm called {\tt SDINT} that combines aspects from several of the aforementioned approaches. Given the extremely extended nature and complexity of our targeted emission, we utilize the most commonly used combination method called feathering, where the combined image is constructed by computing the weighted sum of the single dish and interferometer data in the $uv$-plane. 

We use the {\tt miriad} task, {\tt IMMERGE}, to feather a cube extracted from the Parkes Galactic All-Sky Survey (GASS\footnote{\url{https://www.astro.uni-bonn.de/hisurvey/gass/}}; \citealt{McClure-Griffiths2009}) with our deconvolved ASKAP image. This cube was centered on the SMC and re-gridded to be on the same spectral and spatial grid before combination. {\tt IMMERGE} combines the two data sets by Fourier transforming both the deconvolved ASKAP-only and Parkes images. The Parkes data are then scaled by the ratio of the solid angle of the two beams,
\begin{equation}\label{eq:beamFrac}
\alpha = \frac{\Omega_{\rm ASKAP}}{\Omega_{\rm Parkes}}, 
\end{equation}
where we find $\alpha$ to equal 9.77$\times$10$^{-4}$, assuming the Parkes beam is a Gaussian with a full-width at half maximum (FWHM) of 16$'$ at 1.4 GHz \citep{2010A&A...521A..17K}. This factor accounts for the difference in flux density of the two data sets based solely on the differences in resolution. The ASKAP $uv$ data are then scaled by the factor
\begin{equation}\label{eq:weight}
\beta = 1-\mathscr{F}[B_{\rm Parkes}],
\end{equation}
where the second term represents the 2D Fourier transform ($\mathscr{F}$) of the Parkes beam, $B_{\rm Parkes}$. The two scaled $uv$ data sets are then summed and Fourier transformed back to the image plane. The scaling depicted in Equation~\ref{eq:weight} ensures the effects of the poorly sampled low spatial frequencies in the ASKAP data are smoothly removed before the well-sampled low spatial frequencies provided by the Parkes data are added. The effective beam of the resultant image is the same as the original ASKAP restoring beam. The above scaling assumes the two data sets are perfectly flux calibrated; however, technical issues introduced by stray radiation and spillover make single dish data notoriously difficult to correctly flux calibrate. 

We take an interactive approach to determine the ideal flux calibration factor, henceforth referred to as the single dish factor (SDF), by setting the {\tt factor} keyword in {\tt IMMERGE} to be 0.5, 1.0, and 2.0 and then divide out this same factor in the resulting cube to conserve flux. This wide range of possible SDF values is useful to investigate the extremes of overestimating and underestimating the correction. We then compute the spatial power spectrum (SPS) of several spectral channels. The SPS is defined as
\begin{equation}\label{eq:SPS}
P\left(k\right) = \mathscr{F}\left(k_{\rm RA}, k_{\rm Dec}\right)\times\mathscr{F}^*\left(k_{\rm RA}, k_{\rm Dec}\right), 
\end{equation}
where $k$ is the magnitude ($k=\sqrt{k_{\rm RA}^2 + k_{\rm Dec}^2}$) of the wavenumber ($k_{\rm RA/Dec} = \theta_{\rm RA/Dec}^{-1}$, where $\theta_{\rm RA,Dec}$ is the angle on the sky along RA and Dec in rad),  $\mathscr{F}\left(k_{\rm RA}, k_{\rm Dec}\right)$ is the 2D Fourier transform of the image under study, and the $^*$ symbol represents the complex conjugate. This is a two-point correlation function that measures how power (i.e., structure) is distributed across spatial scales. In practice, the distribution of power is measured by computing the 2D Fourier transforms of the integrated intensity images and measuring the median power in progressively larger annuli. 

\begin{figure*}
\centering{
\includegraphics[width=\textwidth]{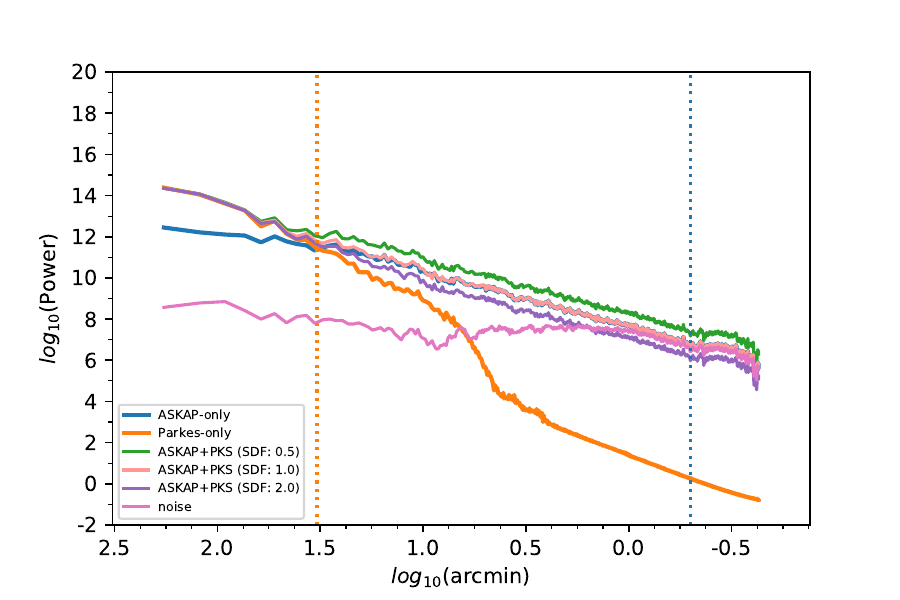}
\caption{\label{fig:SPS_profiles}The SPS profiles of the spectral channel centered on 151.3 km s$^{-1}$ made from an ASKAP-only cube (solid blue), Parkes (solid orange), and several ASKAP+Parkes cubes where the {\tt sdfactor} parameter (referred to as SDF in the text) in {\tt miriad}'s {\tt IMMERGE}, which applies a scale factor to the single dish data before the combination in order to correct for potential flux offsets, is varied. The orange and blue dotted vertical lines denote the maximum recoverable scale of ASKAP based on the smallest baseline distance of 22 m and restoring beam size, respectively. Note that the profile for the SDF=1.0 values lies exactly on top of the ASKAP-only profile towards the smaller angular scales.}}
\end{figure*}

Figure~\ref{fig:SPS_profiles} shows the SPS profiles of the spectral channel centered on 151.3 km s$^{-1}$ in the Local Standard of Rest Kinematic (LSRK; \citealt{Gordon1976}) reference frame produced from combinations that vary the SDF, as well as profiles from the deconvolved ASKAP-only cube and re-gridded Parkes image. Clearly, there is a dearth of power at large scales in the ASKAP-only image, while there is a large decrease in power detected in the Parkes image at scales below the larger single dish beam. All combinations recover the power at large scales since the feather procedure is designed to scale the flux density on those scales to that of the single dish. Interestingly, the variation in SDF has the most notable effect on the power at small scales. When the SDF is underestimated and less than one, the power \textit{increases} at small scales because the scaling to conserve flux after the combination introduces a multiplicative factor that increases the flux of the already existing small-scale structure in the ASKAP-only image. When SDF is overestimated, the power \textit{decreases} at small scales because the low resolution single dish data effectively washes out the observed small-scale structure.

\begin{figure*}
\centering{
\includegraphics[width=\textwidth]{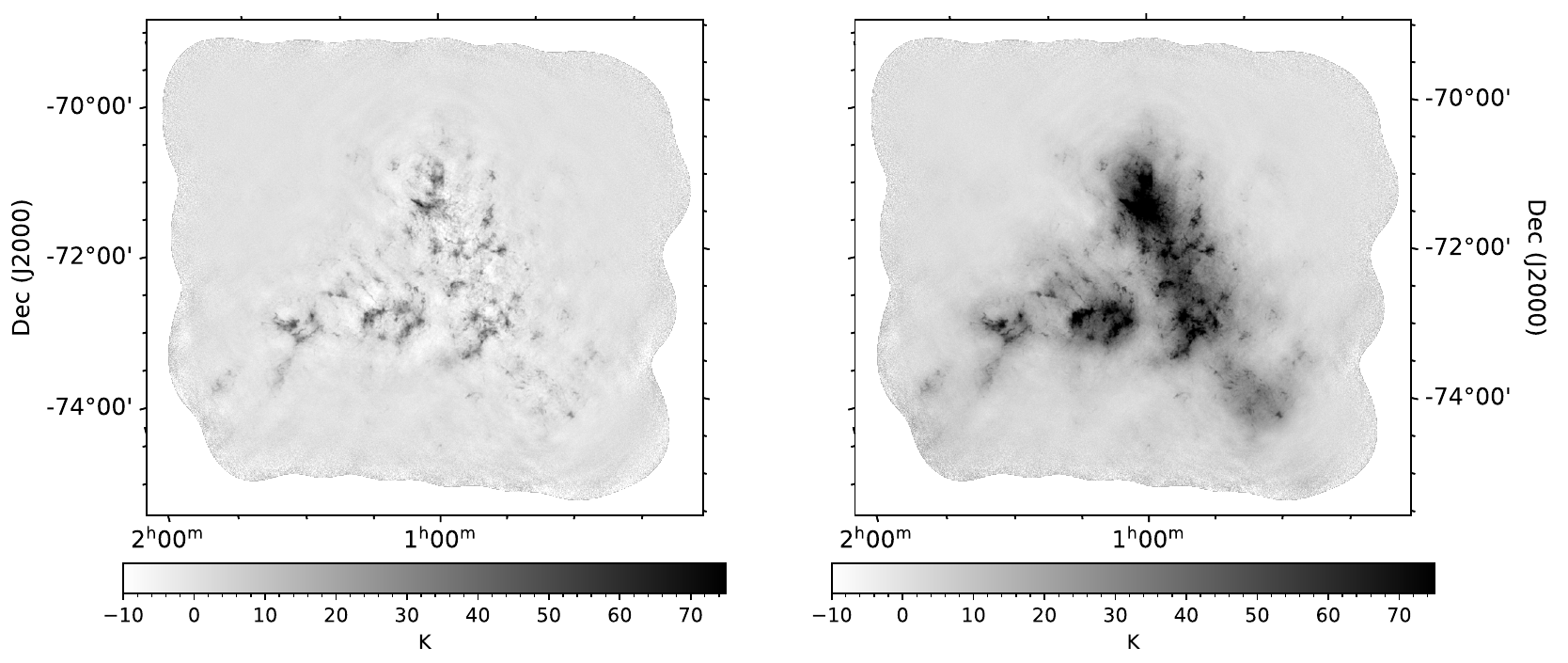}
\caption{\label{fig:combination_comparison} A comparison between the ASKAP-only (left) and resultant combined ASKAP+Parkes image (right) for a single spectral channel centered on 151.3 km s$^{-1}$ in the LSRK reference frame. The optimal SDF factor of 1.0 was applied during the combination.}}
\end{figure*}

We find that an SDF of 1.0 produces virtually an exact agreement with the ASKAP-only profile that extends from the maximum recoverable scale of $\sim$33$'$ down to the size of the 30$''$ restoring beam. Figure~\ref{fig:combination_comparison} shows the negative bowls in the ASKAP-only data, which are caused by the filtering of spatial frequencies corresponding to structures on the largest scales, are completely filled when applying a SDF of 1.0 during the combination. A SDF of 1.0 is reassuring and ensures we do not risk misrepresenting the observed structure by over or under-weighting the power at a given scale. Most importantly, the interesting structures at small scales are not washed-out. SDF values of 0.9 and 1.1 produce noticeable offsets in the SPS profiles, strengthening our conclusion. While we do not show an uncertainty envelope on the SPS profiles for the sake of clarity, Section~\ref{subsec:spatial_frequencies} demonstrates that the typical uncertainties on individual power values are on the order of 5\%.

\subsection{Quality Assessment}\label{subsec:qualityassess}
The spectral line cubes produced from GASKAP-HI observations also need to be validated, both to determine their suitability for science use and also to compare different processing techniques.
We aim to validate the representation of the diffuse \hi\ of the Galactic plane and the Magellanic System in the following ways:
\begin{itemize}
    \item that large scale emission is present in velocity ranges where it is expected and not present where not expected, based on previous surveys;
    \item that the noise level in the emission-free velocity range is close to theoretical, but not below, and is even across the cube; and
    \item that spectra taken at the locations of selected targets, such as against bright sources, do not show any periodic features.
\end{itemize}

We validate the spectral line data at the ASKAP-only stage  rather than after they have been combined with Parkes data.
This allows us to compare the ASKAP-only data with data from GASS; \citealt{McClure-Griffiths2009}) in our tests for expected presence and absence of emission. 
It also provides a more rigorous test of noise levels in the ASKAP data.

To check for the presence or absence of emission, we define two $\Delta v \approx 40~{\rm km~s^{-1}}$ regions where \hi\ line emission is either present or not present in the GASS cube.
We extract a slab from the ASKAP cube covering each velocity region and use {\tt BANE} (part of the {\tt Aegean} suite; \citealt{2018PASA...35...11H}), to produce a large-scale emission line image of each slab.
The large-scale emission line images exclude \hi\ point sources and smooth small-scale fluctuations based on the synthesised beam size.
We assess the maximum flux density level of these images against the thresholds in Table \ref{tab:cube-diagnostics}.

We assess the spectral line noise by taking the standard deviation of each pixel of the off-line slab (as described above) along the spectral axis.
We then compare the median value of this noise map to a fiducial value, the theoretical noise, with the thresholds listed in Table \ref{tab:cube-diagnostics}.
The theoretical noise, at the beam centre and for natural weighting, is calculated as 
\begin{equation}
\sigma_F = \frac{\rm SEFD}{\epsilon_{\rm c} \sqrt{n_{\rm pol} n_{\rm ant} (n_{\rm ant}-1) \Delta_{\rm f} t_{\rm 0}}}
\end{equation}
where SEFD is the System Equivalent Flux Density, $\epsilon_{\rm c}$ is the correlator efficiency, n$_{\rm pol}$ is the number of polarisations, n$_{\rm ant}$ is the number of antennas, $\Delta_{\rm f}$ is the channel width and t$_{\rm 0}$ is the duration of the observation.
ASKAP has a SEFD = 1800 Jy, n$_{\rm pol} = 2$, and n$_{\rm ant} = $ 36 \citep{hotan2021}.
We assume that $\epsilon_{\rm c}$ is included in the SEFD as the SEFD is measured through observations, but it should be $\sim 1$ regardless (see Table~\ref{tab:survey_speed_params}).
A typical GASKAP-HI observation of t$_{\rm 0} = 20$ hours with $\Delta_{\rm f} = 5$ kHz channels gives $\sigma_F = 1.89$ mJy/beam.

\begin{figure}
	\includegraphics[width=\columnwidth]{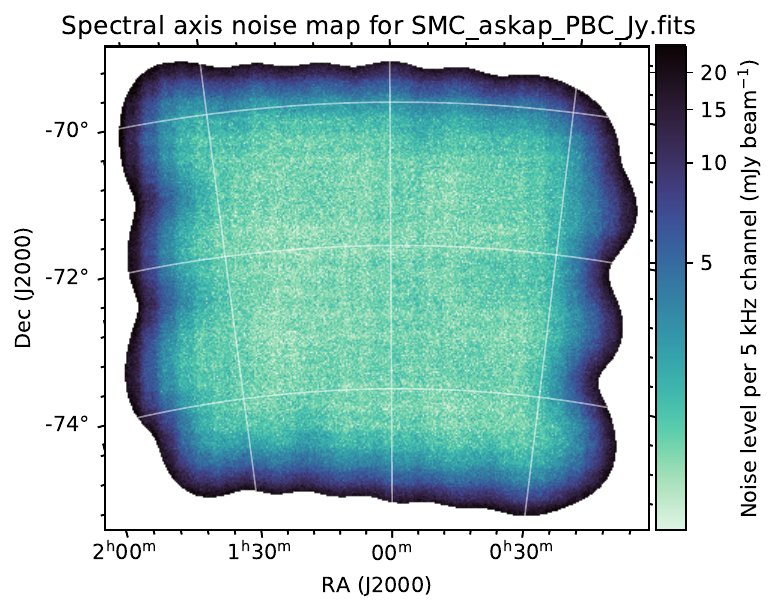}
    \caption{Map of the spectral line noise for the ASKAP-only SMC cube from scheduling blocks 10941 and 10944. 
    There is high noise at the edges of the cube and a notable cross-hatch low level noise pattern in the main area of the cube.
    However, it shows a generally low and consistent noise level across the main area of the cube, with a median noise level of 2.34 mJy beam$^{-1}$, 1.24 times theoretical.}
    \label{fig:spectral_noise}
\end{figure}

\begin{table*}
\centering
\caption{Spectral Line Cube Diagnostics Metrics. 
These are the thresholds for the different quality tests on GASKAP-HI spectral line cubes to assess them with a rating of good, uncertain or bad.}
\label{tab:cube-diagnostics}
\begin{tabular}{lcccl}
\hline
Metric & Good & Uncertain & Bad & Units \\
\hline
On-line emission & $\geq 20$ & $20 - 12$ & $< 12$ & ${\rm Jy~beam^{-1}~km~s^{-1}}$ \\
Off-line emission & $< 5$ & $5 - 12$ & $\geq 12$ & ${\rm Jy~beam^{-1}~km~s^{-1}}$ \\
Spectral Noise compared to theoretical noise & $< \sqrt{2}$ & $\sqrt{2} - 2\sqrt{2}$ & $\geq 2\sqrt{2}$ & -- \\
Spectra with periodic features & 0 & $1 - 4$ & $\geq 5$ & -- \\
\hline
\end{tabular}
\end{table*}

Lastly, we validate spectra extracted both from the locations of the 15 brightest sources (as identified in the continuum catalogue) and the centre of each beam.
At each position we extract a single pixel spectrum and plot these spectra for visual inspection.
We assess the spectra using a partial autocorrelation function (PACF) to identify any periodic features in each spectrum, requiring a 3-sigma result to classify a feature as periodic.
We rate the dataset based on the number of spectra with periodic features as shown in Table \ref{tab:cube-diagnostics}.

We can use these metrics on the ASKAP-only image cube for the SMC field to assess its quality.
Emission is present (21 Jy beam$^{-1}$ km s$^{-1}$) where expected in the range $119 \leq $v$_{\rm LSRK} \leq 165$ km s$^{-1}$ and absent (0.4 Jy beam$^{-1}$ km s$^{-1}$) where not expected in the range $30 \leq $v$_{\rm LSRK} \leq 73$ km s$^{-1}$.
Figure~\ref{fig:spectral_noise} shows a map of the spectral noise across the entire field. The spectral noise is $2.34$ mJy per 5 kHz, or $1.24$ times theoretical, well within the good range. The slightly loss in sensitivty can be attributed to the Briggs weighting applied to the visibilities (see Section~\ref{subsec:noise}.)
The noise is generally smooth across the primary area of the field, however a grid pattern is apparent in the noise at low levels.
This structure is likely a numerical artifact introduced by the Fourier Transform inversion of the visibilities when gridded to a finite grid to return to the image domain during major cleaning iterations.
The effect of this structure is characterized in Section~\ref{subsec:spatial_frequencies}. 
The highest noise is at the edges of the field outside the FWHM of the formed primary beams, as expected.
Finally, there are no periodic features in the test spectra.
With all metrics in the good range we find that the SMC cube is suitable for science use.

\section{Data Products}\label{sec:data_products}

\subsection{Image Cubes}\label{subsec:data_cubes}
The final product of our custom imaging pipeline, after masking the cube at the 10\% level of the combined primary beam response, is a 3901$\times$3471$\times$222 pixel image cube containing jointly deconvolved ASKAP data feathered with regridded Parkes data that covers a LSRK velocity range of 40.0 km s$^{-1}$ to 253.9 km s$^{-1}$ at a resolution of 0.98 km s$^{-1}$. This cube is publicly accessible through a unique Digital Object Identifier (DOI)\footnote{\url{https://doi.org/10.25919/www0-4p48}}. We measure the rms noise using the source finding software, {\tt SoFiA 2; \citealt{serra2015, westmeier2021}}, which fits a Gaussian to the negative half of the histogram of pixel values. This method is more robust against the presence of strong, extended emission or artifacts than measuring the standard deviation of emission-free channels. The width of this Gaussian is 1.1 K at the restoring beam size of 30$''\times$30$''$, corresponding to a 5$\sigma$ $\hi$ column density detection limit of 4.4$\times$10$^{19}$ cm$^{-2}$ over a 20 km s$^{-1}$ linewidth. This cube is the most sensitive and highest resolution image (30$''\times30''$) of the SMC in $\hi$ ever achieved.

\subsection{Noise as a Function of Angular Resolution}\label{subsec:noise}
Two key science targets of GASKAP-HI are diffuse $\hi$ emission distributed on angular sizes larger than the synthesised beam and cold gas as revealed through sensitive $\hi$ absorption towards background continuum sources. It is critical to ensure the brightness temperature rms noise, $\sigma_{\rm T}$, is minimized and behaves as predicted at the angular resolutions relevant for key GASKAP-HI science. 
\citet{johnston2007} give the survey speed ($SS$) of ASKAP in terms of $\sigma_{\rm T}$ as
\begin{equation}\label{eq:survey_speed}
    SS = FBn_{\rm pol}\left(\frac{\epsilon_{\rm c}\sigma_{\rm T}}{T_{\rm sys}}\right)^2\left(\frac{f}{\epsilon_{\rm s}}\right)^2,
\end{equation}
where $B$ is the spectral resolution, $F$ is the FoV, $n_{\rm pol}$ is the number of polarizations (2), $\epsilon_{\rm c}$ is the correlator efficiency, $T_{\rm sys}$ is the system temperature in emission-free regions, and the filling factor of the array
\begin{equation}\label{eq:covering_factor}
    f = \frac{A\eta N\Omega\epsilon_{\rm s}}{\lambda},
\end{equation}
which describes the covering factor of the array in the aperture plane. Here, $A$ is the area of a single antenna, $\eta$ is the aperture efficiency for a single antenna, $N$ is the total number of antennas, and the solid angle of the synthesized beam $\Omega$ = 1.13($\theta_{\rm s})^2$ in steradians (sr), where the size of the synthesized beam is characterized by the FWHM along the major and minor axis: $\theta_{\rm s}=\sqrt{\theta_{\rm maj}\cdot\theta_{\rm min}}$

Finally, $\epsilon_{\rm s}$ is what is known as the synthesized aperture efficiency, constrained to be $\leq$~1 and relates to the weighting of the gridded visibilities in the $uv$-plane before inversion to the image plane. For example, applying a Briggs weighting scheme reduces $\epsilon_{\rm s}$, since some baselines in the $uv$-plane become higher weighted than others. Keeping all other parameters in Equations~\ref{eq:survey_speed} and \ref{eq:covering_factor} fixed, we anticipate the rms brightness temperature to scale as $\sigma_{\rm T}$ $\propto f^{-1} \propto \theta_{\rm s}^{-2}$, where $\epsilon_{\rm s}$ encodes the removal of baselines from the spatial smoothing. The telescope specifications are summarized in Table~\ref{tab:survey_speed_params}.

\begin{table*}
\centering
\caption{Noise Scaling Parameters}
\label{tab:survey_speed_params}
\begin{tabular}{lcc}
\hline
Parameter & Symbol & Value [units] \\
\hline
Field-of-view & $F$ & 25 [deg$^{2}$] \\
Dish Area & $A$ & 113 [m$^{2}$] \\
Spectral Resolution & $B$ & 5 [kHz] \\
Number of polarisations & $n_{\rm p}$ & 2 \\
Correlator efficiency & $\epsilon_{\rm c}$ & $\sim$1 \\
Number of antennas & $N$ & 36 \\
Aperture efficiency & $\eta$ & 0.7 \\
(Emission-free) system temperature & $T_{\rm sys}$ & 55 [K] \\ 
rms noise & $\sigma_{\rm T}$ & 1.1 [K] \\ 
Dwell Times & $t_{\rm int}$ & 12.5, 20.9 50, 200 [h] \\
\hline
\end{tabular}
\end{table*}

\begin{figure}
\centering{
\includegraphics[width=\columnwidth]{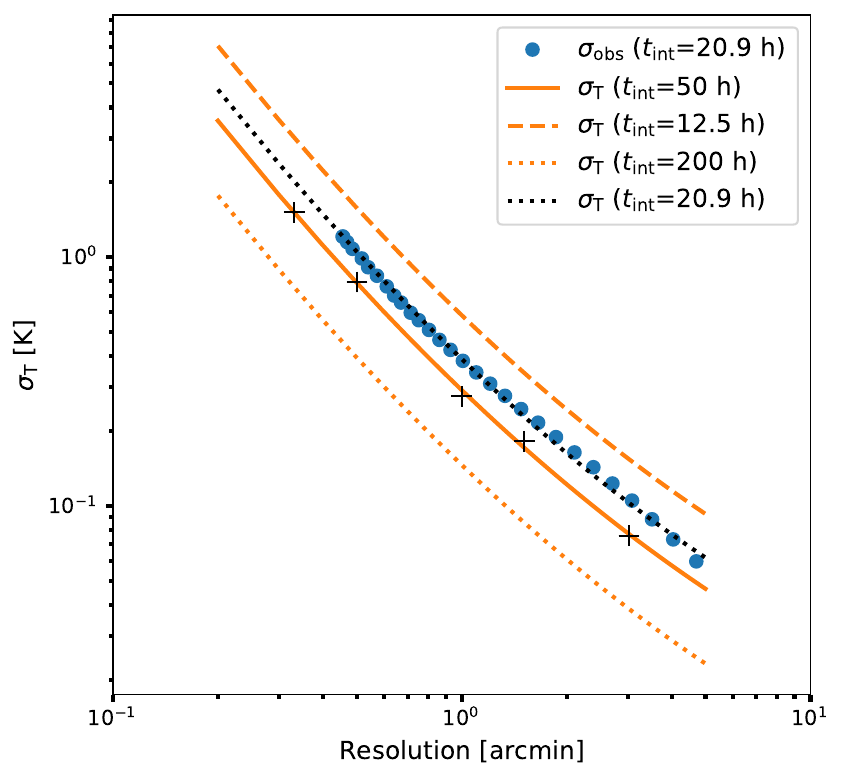}
\caption{\label{fig:noise_profiles}The measured noise (blue points) within a sub-region of our GASKAP-HI cube after applying gradually larger tapers to the baselines in the $uv$-plane compared with the predictions shown in Figure 4 and Table 3 from \citet{dickey2013} (orange lines and black crosses). The dotted black line denotes the updated theoretical noise using the measured $T_{\rm sys}/\eta$ scaled to an integration time of 20.9 h.}}
\end{figure}

Figure~\ref{fig:noise_profiles} compares the measured noise as a function of angular resolution with the predictions from \citet{dickey2013} (orange lines and crosses) at varying angular resolutions. The predicted noise profiles are determined by solving Equation~\ref{eq:survey_speed} for $\sigma_{\rm T}$ for various dwell times ($t_{\rm int}$) by setting $SS$=$F$/$t_{\rm int}$. We measure the noise in our cube by imaging a sub-region of 4 beams $\times$ 3 interleaves of 20 contiguous emission-free channels with natural weighting. We then apply a Gaussian taper to the baselines in the $uv$-plane before the inversion to the image plane to alter the final spatial resolution. The final beam size is a fit to the main lobe of the resulting PSF after tapering. We create 30 sub-cubes each with a different taper size that ranges from 12,000$\lambda$ to 400$\lambda$ equally spaced in log-space. The final noise is again taken to be the width of a Gaussian fit to the negative half of the histogram of pixel value as determined by {\tt SoFiA}.

The measured noise is in excellent agreement with the theoretical predictions scaled to match the current telescope specifications (dotted black line). The profiles begin to deviate from the predicted $-$2 power-law at larger resolutions due to the dependence of $\epsilon_{\rm s}$ at larger $uv$ tapering. The agreement between the expected and measured scaling of the noise is confirmation that the data products of GASKAP-HI are indeed pioneering in terms of the trade-off between rms brightness temperature with angular resolution for the study of nearby $\hi$. 

\subsection{Represented Spatial Frequencies}\label{subsec:spatial_frequencies}
We further characterize the range of spatial frequencies represented in our final images after multi-scale joint deconvolution by employing the SPS as a data quality assessment tool. Specifically, we investigate whether a two-dimensional Gaussian function with a FWHM of 30$''$ ($\sim$ 8 pc at the distance of the SMC) is a good approximation for the restoring beam across all available scales in the final image. 

Similar to Section~\ref{subsubsec:combination}, we calculate the two-dimensional angular power spectrum for each spectral plane of our final image cube. Many studies that apply the SPS to column density or intensity images of various observational tracers in the Milky Way show the density distribution of the ISM as a function of scale is well described with a single power law (e.g., \citealt{Martin2015, miville-deschenes2016, pingel18a}). We therefore fit the angular power spectrum profile $P\left(k\right)$ in each velocity channel with a model
\begin{equation}\label{eq:singlePL_model}
P\left(k\right) = B\left(k\right)Ak^{-\gamma} + QP_{\rm noise}\left(k\right), 
\end{equation}
where $B\left(k\right)$ is the power spectrum of our ideal 30$''$ restoring Gaussian beam to account for instrumental systematics that affect the overall shape of the SPS,~$A k^{-\gamma}$ is the ISM power-law term, and $Q$ is a multiplicative factor to the noise floor. The noise template, $P_{\rm noise}\left(k\right)$, is estimated by taking the mean SPS profile for 40 emission-free channels. After fitting for $A$, $\gamma$, and $Q$ in each velocity channel, Equation~\ref{eq:singlePL_model} can be rearranged to solve for 
\begin{equation}\label{eq:transfer_function}
\widetilde{B}\left(k, v_{\rm LSRK}\right) = \frac{P\left(k, v_{\rm LSRK}\right)-Q\left(v_{\rm LSRK}\right)P_{\rm noise}\left(k\right)}{A\left(v_{\rm LSRK}\right) k^{-\gamma\left(v_{\rm LSRK}\right)}},
\end{equation}
where $A$, $Q$, and $P\left(k\right)$ now depend on $v_{\rm LSRK}$, the observed radial velocity in the LSRK reference frame; $\widetilde{B}\left(k, v_{\rm LSRK}\right)$ is the global transfer function, which similar to optical systems specifies how the true sky brightness distribution is represented in our image cube. For example, the absence of structure in the transfer function indicates our data is the true sky brightness distribution convolved with our 30$''$ Gaussian restoring with added radiometer noise. 

In order to fully characterize the uncertainty on our model parameters, we follow similar fitting procedures from \citet{koch2020}, who used the {\tt PYMC3} package \citep{salvatier2015}, to run a Markov Chain Monte Carlo (MCMC) sampling to fit our final models. Due to the large range of angular scales spanned by the $\hi$ emission, we assign uniform priors for each of our model parameters and sample the amplitude $A$ in $log_{10}$ space:
\begin{equation}\label{eq:A_prior}
log_{10}A \sim \mathcal{U}(-20, 20)
\end{equation}
\begin{equation}\label{eq:gamma_prior}
\gamma \sim \mathcal{U}(0, 5)
\end{equation}
\begin{equation}\label{eq:Q_prior}
Q \sim \mathcal{U}(0, 10)    
\end{equation}
We found that the fitting routine converges quickly with these choices of priors and the final fits are not affected by reasonable changes to the range of these priors. The MAD is used as an estimate of the uncertainty in each point $P_i$ in the azimuthally-averaged profile
\begin{equation}\label{equation}
    \sigma_{P(k)} = \mathrm{median}\left(\left|P_i - \mathrm{median}\left(P_{\rm annuli}(k)\right)\right|\right).
\end{equation}
We compute the absolute deviation from the median power for each pixel within each annulus in the spatial frequency domain and take the final uncertainty to be the median of these deviations. The MAD is more robust against bias from small sample sizes at the smallest annuli and outlying values than the standard deviation. Each value of $P(k)$ is treated as an independent sample drawn from a normal distribution with width equal to the MAD values in each azimuthal bin. These samples are first drawn in $log_{10}$ space then converted to linear scaling to avoid negative values for the power spectrum. Finally, we restrict our fits to a range of spatial frequencies between $k < k_{\rm min}/3$ ---  where $k_{\rm min}$ is one half the inverse of the largest map dimension --- and $3k_{\rm max}$, where $k_{\rm max}$ corresponds to the Gaussian standard deviation of our 30$''$ ideal beam. Fitting this range of spatial frequencies avoids the fluctuations from the large-scale power and bias from small-scale statistics at the smallest annuli, while also removing influence from correlated small-scale structure below the size of the restoring beam. 

\begin{figure*}
    \centering
        \includegraphics[width=0.9\textwidth]{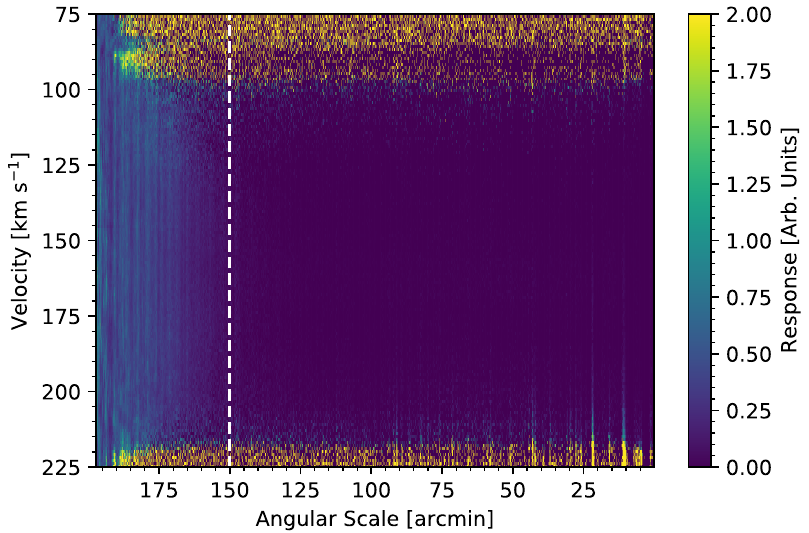} 
        \includegraphics[width=0.9\textwidth]{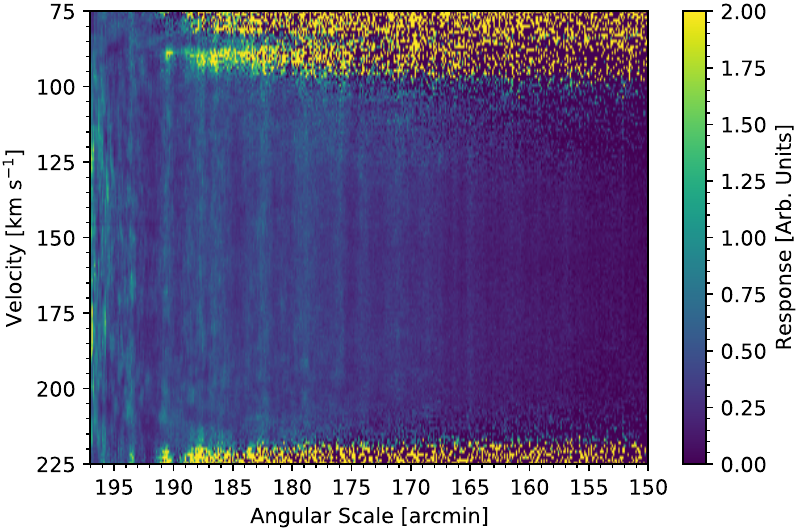} 
\caption{Global transfer function of the SMC image cube. The transfer function is useful for determining whether the observed sky brightness distribution is the true sky brightness distribution convolved with our 30$''$ Gaussian restoring beam with added radiometer noise. The top panel shows the entire range of angular scales, while the bottom panel shows the response between 200$'$ to the vertical dashed line at 150$'$ in the top panel to highlight scales with the most significant structure in the transfer function. The pixelated regions denote mostly emission-free channels.\label{fig:tf_summary}}
\end{figure*}

Figure~\ref{fig:tf_summary} shows the measured transfer function for our SMC image cube. The absence of structure in the transfer function at small angular scales indicates that the density distribution of $\hi$ in the SMC is well described by a single power law down to the resolution limit of 10 pc, consistent with the conclusions from the power spectrum analysis of \citet{stanimirovic1999}, which probed $\hi$ in a combined ATCA and Parkes image cube down to 30 pc. We discuss the interpretation of the $\hi$ turbulent properties in Section~\ref{subsec:turb_properties}

While the small-scale structure is well represented by the ISM power-law, there is some structure towards the largest angular scales ($\gtrsim$ 160$'$), highlighted in a zoomed version in the bottom panel of Figure~\ref{fig:tf_summary}. There are several possible explanations for this discrepancy including: the presence of large scale deconvolution residuals, imperfect combination with the Parkes data, and large-scale structure in the noise. In regards to the presence of structure in the noise, the noise template SPS profile shown in the top panel of Figure~\ref{fig:SPS_profiles} does begin to slightly deviate from the flat profile expected for purely random Gaussian distributed noise at these larger angular scales and also roughly correlates with the size scales of the low-level numerical artifacts from the Fourier transforms seen in Figure~\ref{fig:spectral_noise}. Consequently, the structure in the transfer function at the largest angular scales likely manifests from a combination of sources of error, including bias from small-number statistics in the azimuthal binning of the smallest annuli and the known structure in the noise. It could also trace the azimuthally averaged baseline distribution. However, the power of these artifacts in the noise is many orders of magnitude below the power of the signal at these spatial frequencies and ultimately can be accounted for by restricting the range of scales (e.g., up to $\sim$150$'$) fit during a power spectrum analysis.

\begin{figure*}
    \centering
        \includegraphics{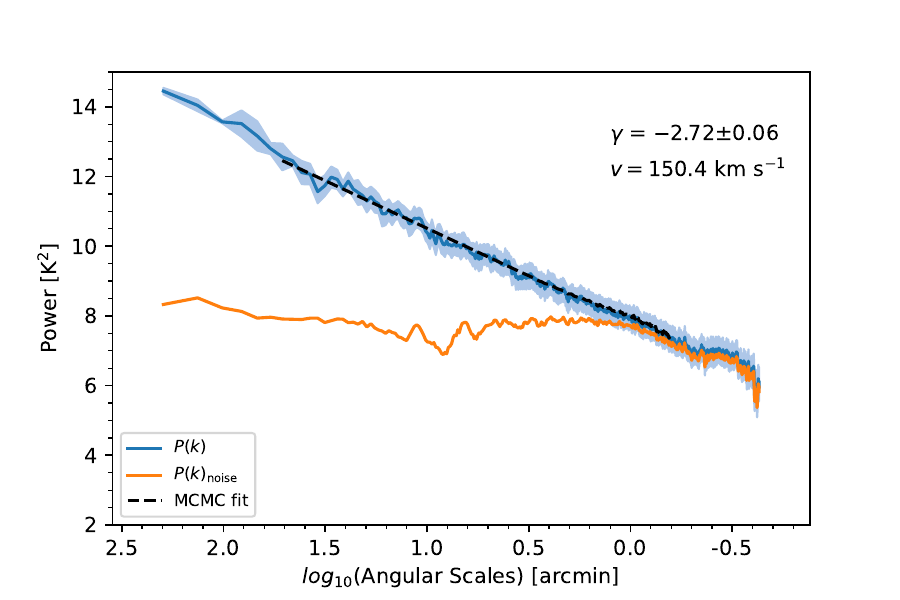} 
        \includegraphics{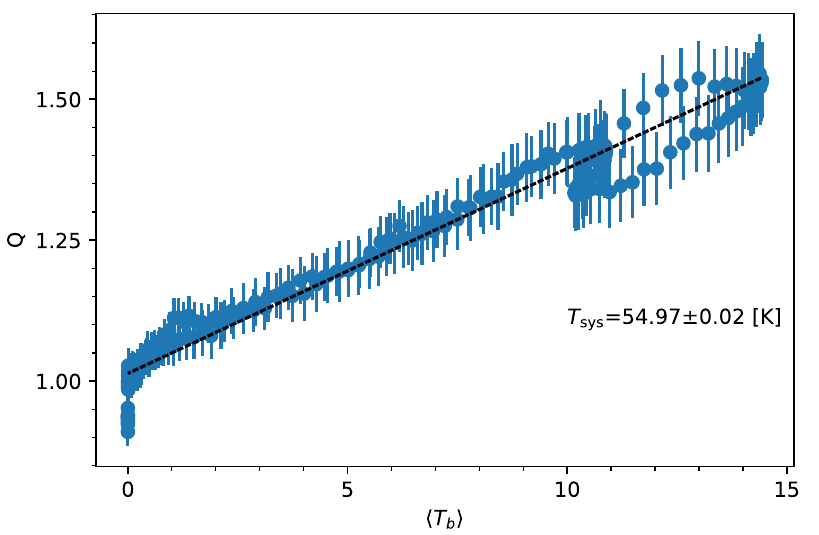} 
\caption{Top: an example fit of our single component power-law to a spatial power spectrum profile from a representative emission channel with the uncertainty derived from the MAD metric denoted by the shaded region. The noise template SPS is shown in orange. Bottom: $Q$, the multiplicative factor applied to the template SPS noise profile, as a function of the mean $T_{\rm b}$ of each spectral channel. A linear fit to channels with significant emission returns the expected $T_{\rm sys}$ of ASKAP at 1.4 GHz.  \label{fig:sps_fit_summary}}
\end{figure*}

For example, the emission SPS profile from the same representative spectral channel in the top panel of Figure~\ref{fig:sps_fit_summary} is exceptionally well-fit by our model. Inspection of other channels with extended emission show similar convergence. By restricting the range of angular scales and allowing a multiplicative factor for the noise, we ensure the presence of artifacts is taken into account in our SPS analysis. \citet{Martin2015} demonstrated through similar SPS that the noise is effectively the combination of receiver and sky noise \citep{boothroyd2011}, $Q = 1 + T_{\rm 0, b}/T_{\rm sys}$, where $T_{\rm 0, b}$ is a characteristic brightness temperature. We slightly alter the form of this relationship to $Q = 1 + 2 T_{\rm 0, b}/T_{\rm sys}$, where the factor of 2 accounts for the differences in the definitions of the linear XX and YY polarizations between \citet{boothroyd2011} and the {\tt ASKAPsoft} calibration pipeline, i.e. (XX+YY) versus (XX+YY)/2. In the bottom panel of Figure~\ref{fig:sps_fit_summary}, we plot the fitted $Q$ values versus the mean brightness temperature in each spectral channel. A fit of a simple linear function to channels with significant emission returns 
$Q = 1.014(\pm0.003) + 0.0182(\pm0.0004)T_{\rm 0, b}$, which gives $T_{\rm sys}$ = 54.97$\pm$0.02 K. This agrees exceptionally well with the expected $T_{\rm sys}$ at 1.4 GHz \citep{hotan2021}.   

\section{Gas Morphology}\label{sec:hvc}
\subsection{General Properties}\label{subsec:gen_properties}

\begin{figure*}
	\includegraphics[width=\textwidth]{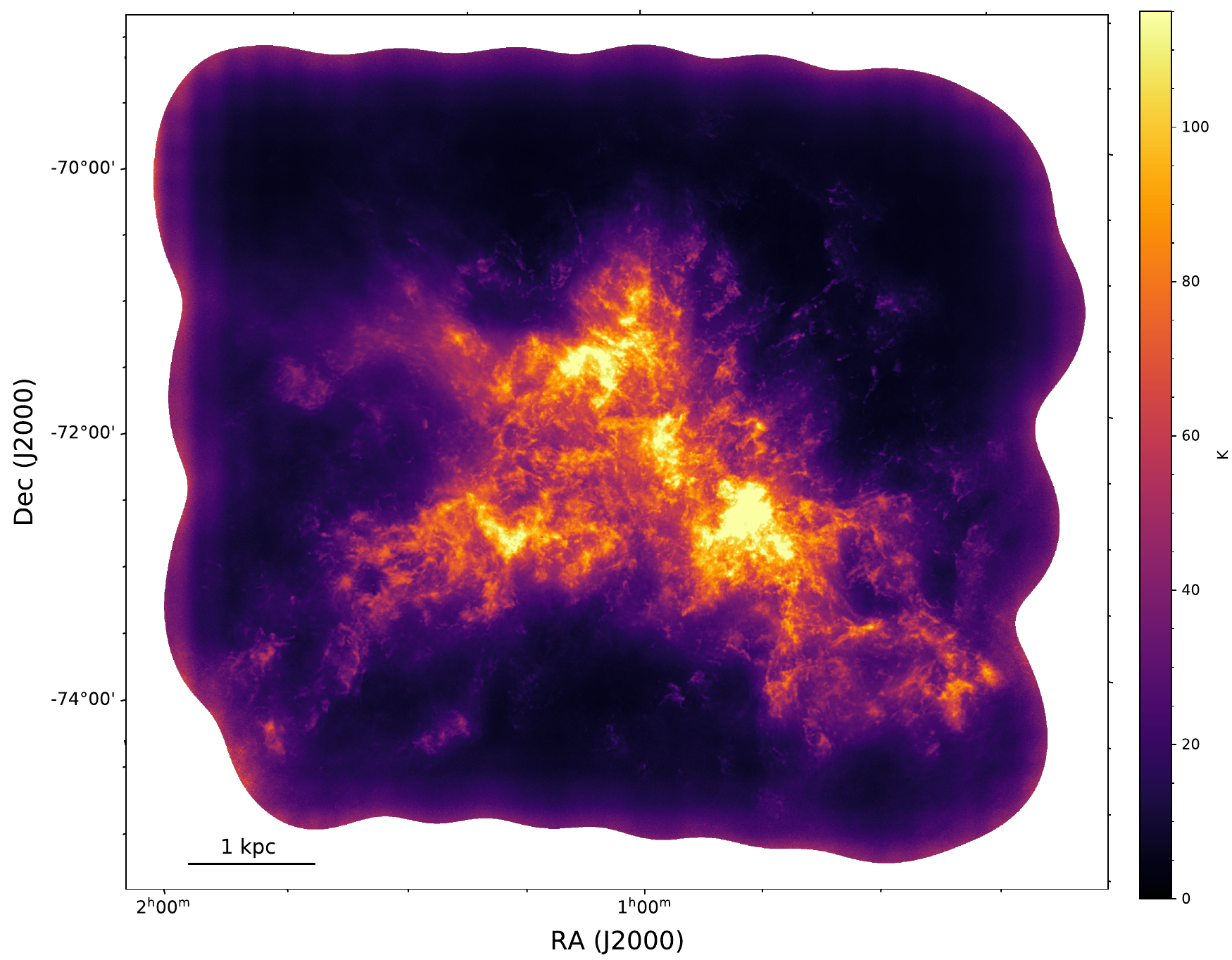}
    \caption{The peak intensity along each line-of-sight from our combined ASKAP and Parkes image cube of the SMC.}
    \label{fig:peak_intensity}
\end{figure*}

\begin{figure*}
	\includegraphics[width=\textwidth]{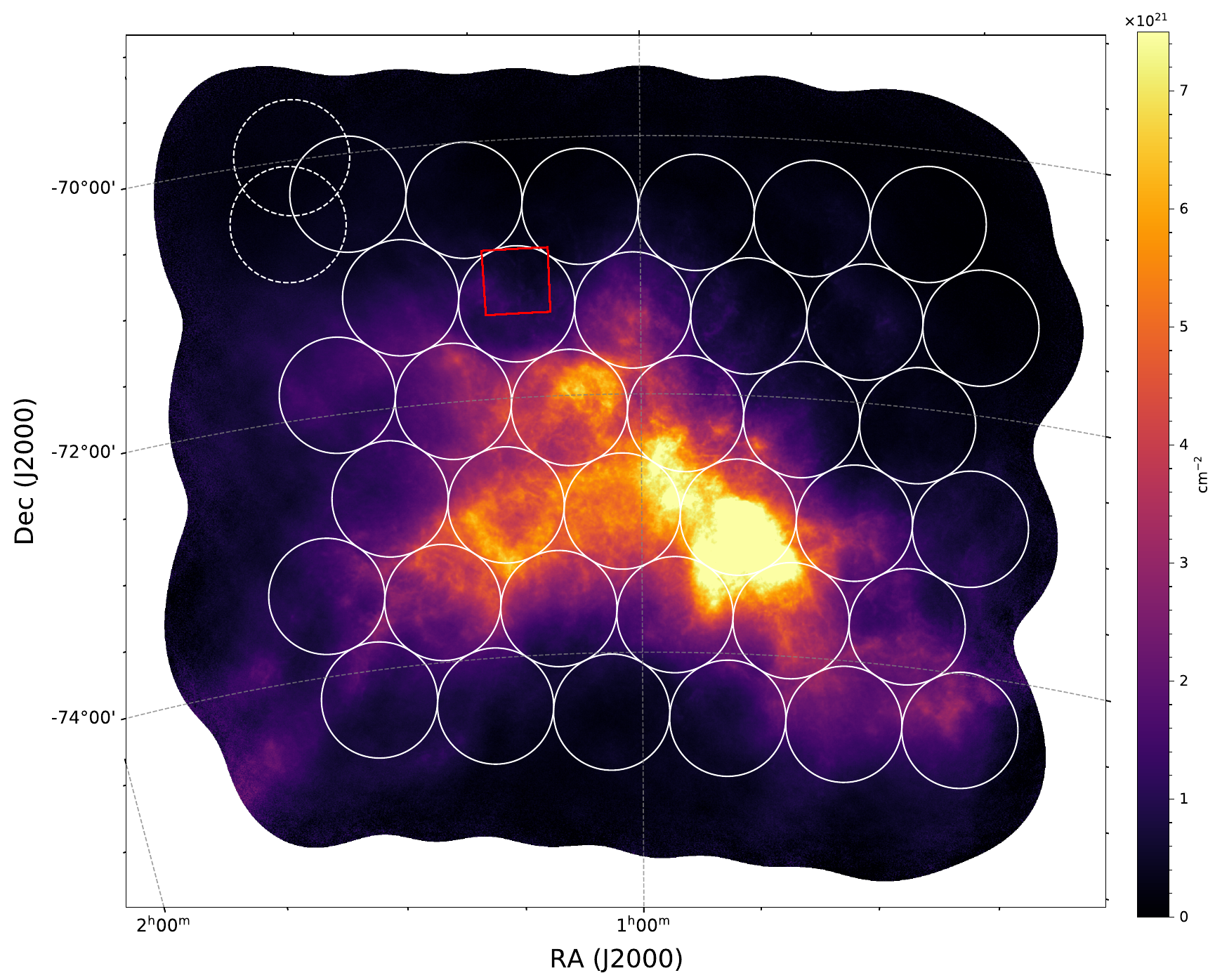}
    \caption{ASKAP+Parkes $\hi$ column density image of the SMC produced from Equation~\ref{eq:column_density} integrated over the LSRK velocity range 60.5 km s$^{-1}$ to 235.4 km s$^{-1}$. The solid white circles denote the locations of the 36 formed beams from interleave A. In order to demonstrate the positional offset of the PAF footprint for each interleave position, we overlay dashed circles at the locations of beam 36 in interleave B (upper) and C (lower). This interleaving scheme ensures even sensitivity, and thus a consistent noise level across the entire instantaneous 25 deg$^2$ FoV. The red rectangle region contains the high velocity cloud discussed in Section~\ref{subsec:hvc}.}
    \label{fig:column_density}
\end{figure*}

The peak $\hi$ intensity image presented in Figure~\ref{fig:peak_intensity} demonstrates the full extent of the emission captured within a single $\sim$25 deg$^2$ ASKAP footprint. While not a true representation of the projected $\hi$ density distribution, as we simply select the peak brightness temperature value along the spectral profile at each spatial pixel, this spectacular image reveals that an incredible wealth of small-scale structure permeates the overall smooth diffuse $\hi$ component of the SMC. The extensive exo-galactic population of $\hi$ features identified in \citet{McClure-Griffiths2018} are clearly visible throughout the northern regions of this map. We further highlight the projected density distribution by computing the column density image shown in Figure~\ref{fig:column_density}, assuming the $\hi$ is optically thin:
\begin{equation}\label{eq:column_density}
    N(x,y)_{\rm HI} = 1.82\times10^{18} \int T_{\rm b}\left(x,y, v\right) dv \;{\rm cm^{-2}}, 
\end{equation}
where the brightness temperature, $T_{\rm b}\left(x,y\right)$, at each spatial pixel is integrated over the spectral range 235.4 km s$^{-1}$ to 60.5 km s$^{-1}$. The boundaries of the ASKAP primary beams are overlaid to demonstrate that the size scales of emission range from a single synthesized beam element up to the extent of multiple formed primary beams. Strikingly, there are still some indications of complex filamentary structure in the column density image, highlighting that our increased angular resolution resolves a wealth of new cold narrow features even when integrating across the entire spectral range. 

\begin{figure*}
	\includegraphics[width=0.9\textwidth]{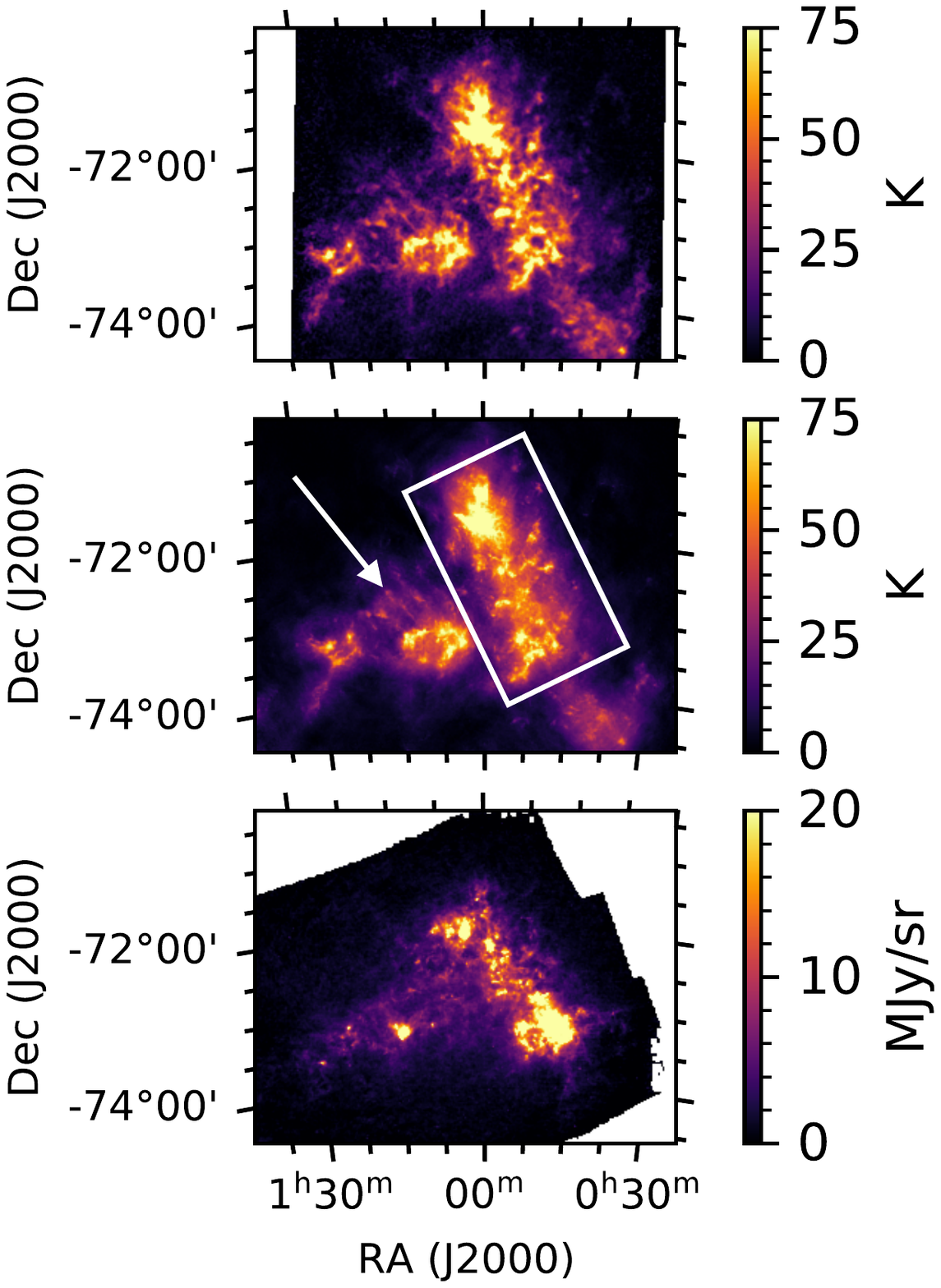}
    \caption{A visual comparison between a single spectral channel centered on 151.3 km s$^{-1}$ in the LSRK reference frame from the previous generation ATCA+Parkes of the SMC in $\hi$ (top; \citealt{stanimirovic1999}), the new GASKAP-HI (ASKAP+Parkes) $\hi$ image (middle), and the 250$\mu$m dust image from the Herschel space observatory (bottom; \citealt{gordon2014}). The white rectangle outlines the Bar region of the SMC and the white arrow points to resolved $\hi$ plumes in the Wing region. The extent and detail of the small-scale $\hi$ features in this individual spectral channel of the GASKAP-HI image are similar to those seen in the dust continuum emission.}
    \label{fig:hi_vs_dust}
\end{figure*}

The single channel shown in the top two panels of Figure~\ref{fig:hi_vs_dust} highlights the improvement in angular resolution in our pilot data over the previous generation image made with ATCA and Parkes. It could be argued that intermediate scale features appear better recovered in the previous generation ATCA+Parkes image. This is a combined effect of including emission over a slightly larger spectral range (1.65 km s$^{-1}$), spatial smoothing to a lower angular resolution, and the difference in the $uv$-weighting applied during imaging. \citet{staveley-smith1997} applied a robustness parameter of 0.0, giving slightly less weight to sparsely sampled (i.e., long) baselines, bringing out structure on these intermediate scales.

The abundance of small-scale $\hi$ features in the Bar of the SMC, extending from the north to the south-west and outlined by the white rectangle, trace the stellar feedback effects from the strong ongoing star formation in this region as seen in H$\alpha$ \citep{winkler2015} and other star burst regions such as the N66/NGC 346 complex \citep{heydari2010}. The detail in small-scale features in a single spectral channel from our new GASKAP-HI data approaches those traced by the far-infrared 250$\mu$m dust image produced as part of the HERschel Inventory of The Agents of Galaxy Evolution (HERITAGE) project \citep{meixner2013}. A comparison of the properties of the $\hi$ from a single spectral channel with dust emission is not straightforward given the continuum nature of the re-radiated emission arising from the cool dust. Nevertheless, assuming the gas and dust are well-mixed due to the correlation between color excess and total hydrogen column density \citep{rachford2009}, future studies will utilize the improved angular resolution to provide insights into spatial variation of the dust-to-gas ratio across the entirety of the SMC. 

Towards the so-called Wing of the SMC, which extends towards the LMC, the superior spectral and spatial resolution of these data reveal a complex network of discrete linear plumes (see the arrow in the middle panel of Figure~\ref{fig:hi_vs_dust}) with projected lengths of $\sim$ 0.5 kpc. A similar plume stemming just east of the northern portion of the Bar extends almost 1 kpc. The spectral resolution of our new image cube ensures these plumes and other interesting features, such as the supershell centered on RA = 00$^{\rm h}$50$^{\rm m}$, Dec=$-$73$^{\rm d}$10$^{m}$, are fully resolved over 5 to 20 channels. Section~\ref{subsec:hvc} demonstrates how our improved insight into the kinematic properties of these discrete features helps with constraining important physical properties of the SMC. 

\begin{figure*}
\begin{center}
\includegraphics[width=0.47\textwidth]{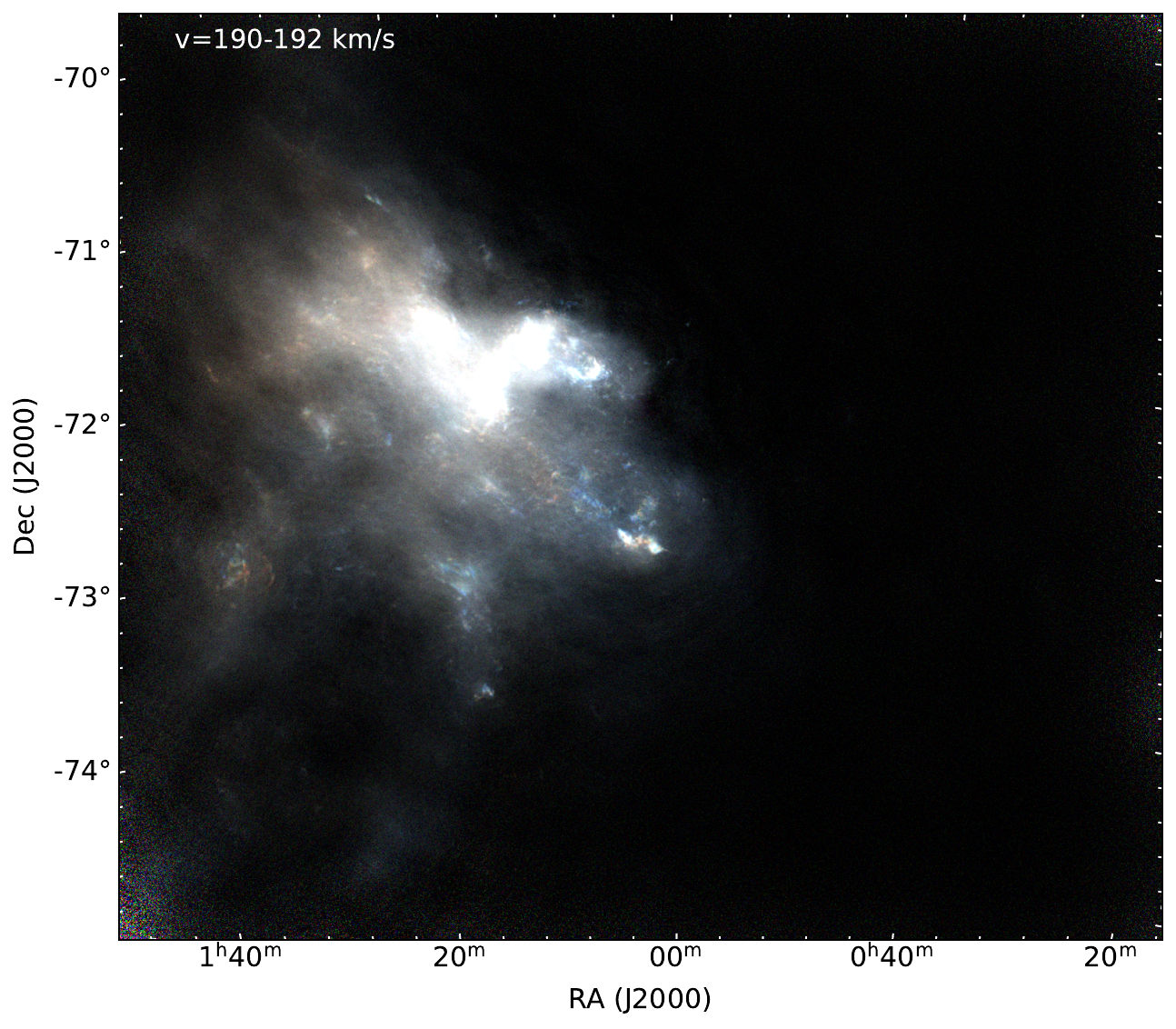}
\includegraphics[width=0.47\textwidth]{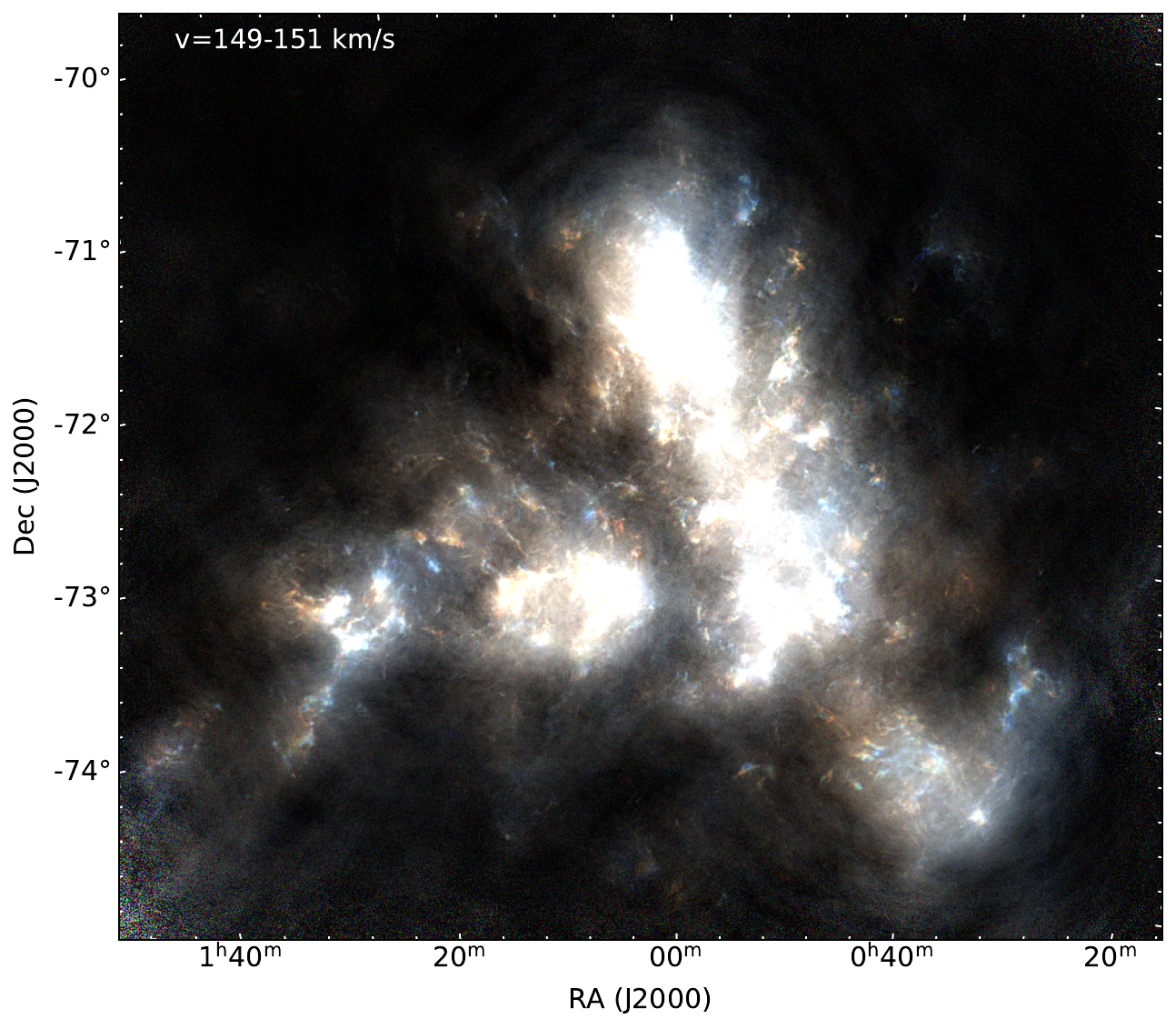}
\includegraphics[width=0.47\textwidth]{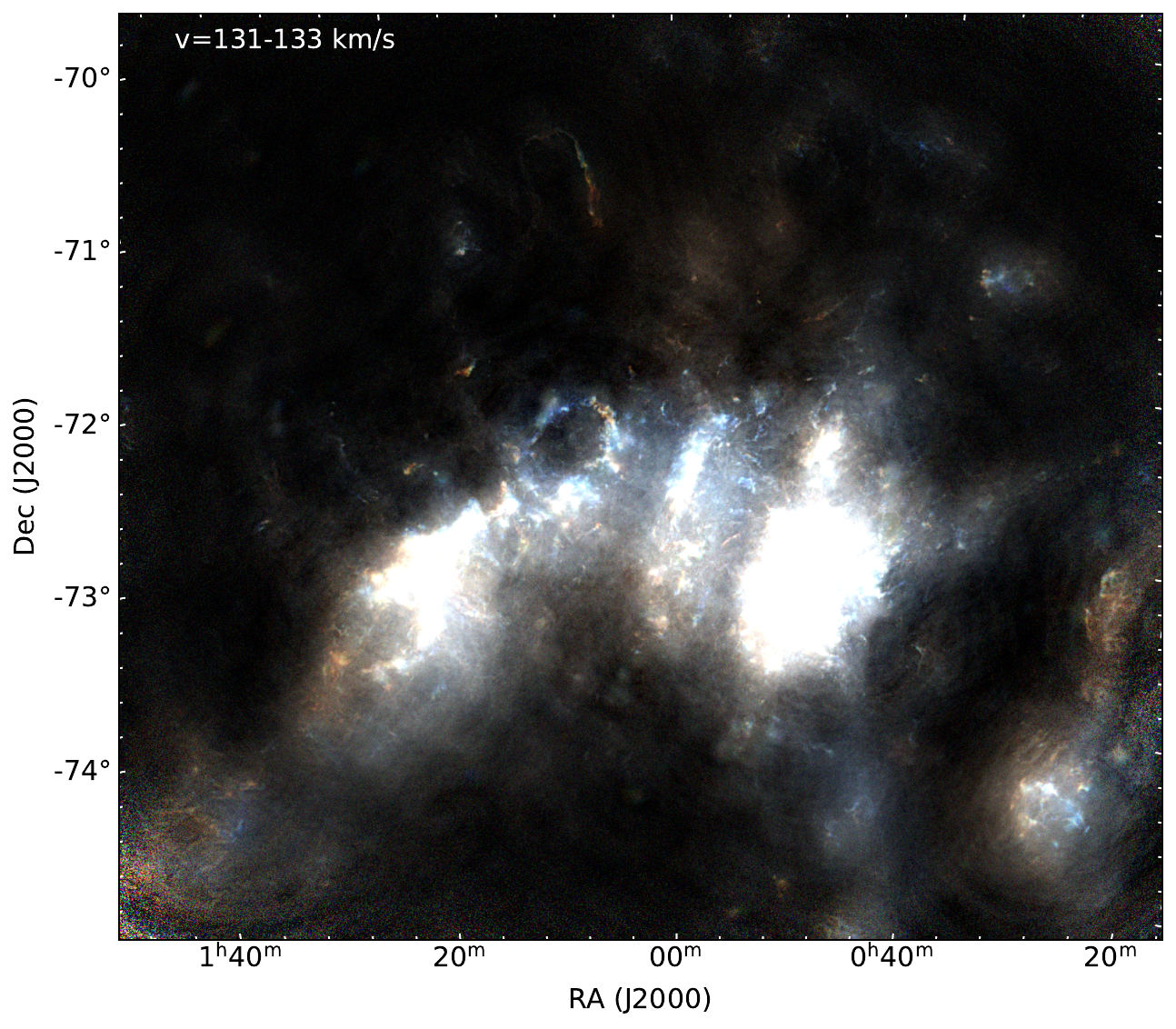}
\includegraphics[width=0.47\textwidth]{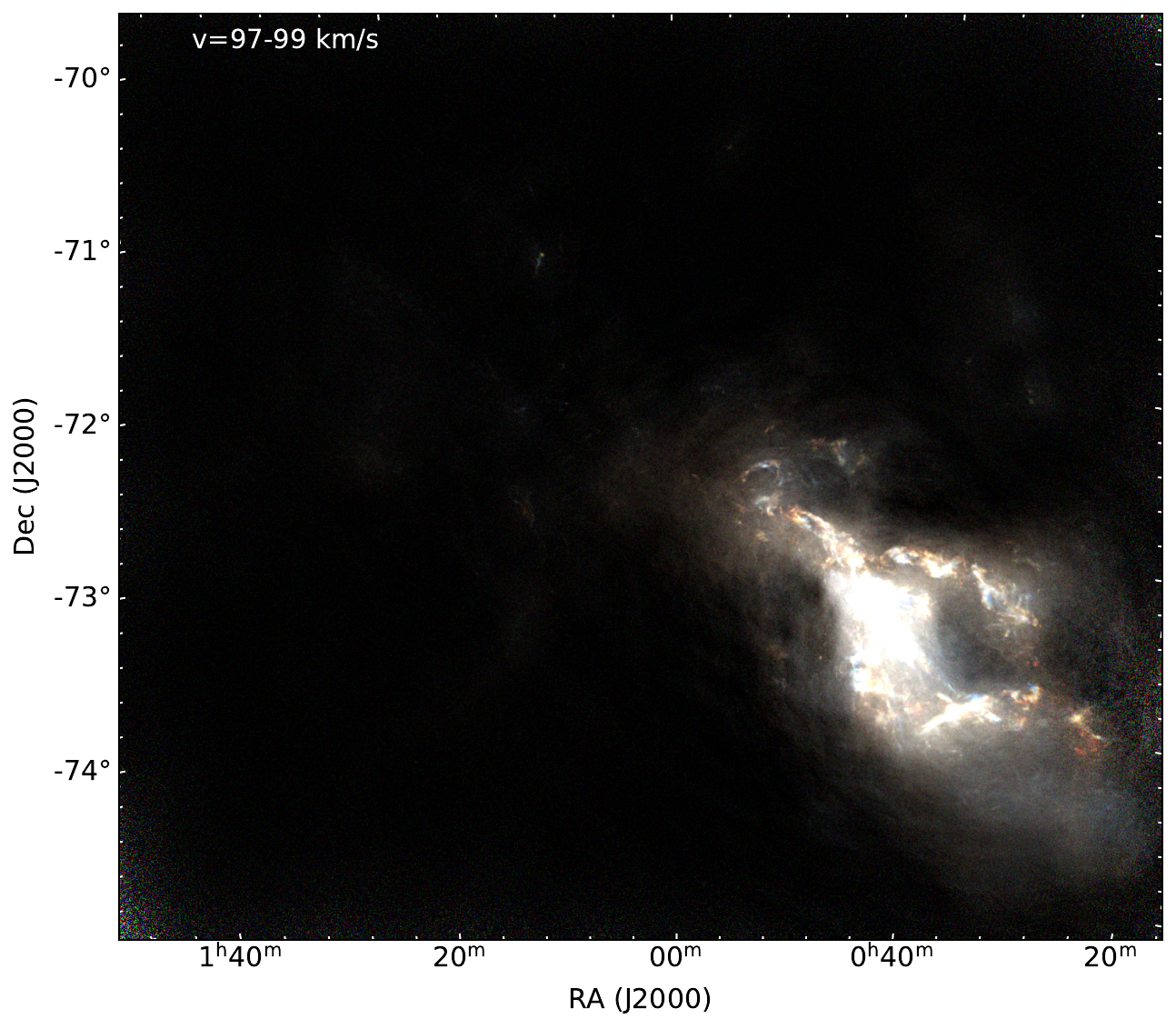}
\end{center}
    \caption{Selected channel maps of the combined ASKAP+Parkes $\hi$ image cube. Red, green and blue are assigned sequentially to adjacent velocity channels, each displayed with an arcsin colormap scaling over a brightness temperature range of $T_{\rm b}$ = 1 K and 70 K. The colorscale of all panels is the same. The LSRK velocity range is listed in each panel.}
    \label{fig:chan_maps}
\end{figure*}

The three-colour images of selected spectral channel maps in Figure~\ref{fig:chan_maps}, which stack emission from adjacent channels, highlight the complexity and narrow spectral properties of the newly revealed small-scale structure. At more redshifted velocities relative to the systemic velocity of 148 km s$^{-1}$ LSRK \citep{diTeodoro2019b}, the $\hi$ morphology is relatively smooth at large scales but gradually transitions to a filamentary form on arcminute scales as the velocity decreases. The $\hi$ distribution becomes increasingly intricate near the systemic velocity, taking the form of plumes, filaments, knots, and shells. The discrete filamentary features are analogous to the slender $\hi$ fibres shown by \citet{clark2014} to be aligned with the interstellar magnetic field of the Milky Way, while the small-scale knots and arcs trace structure on 10 pc scales, similar to the typical size of star-forming regions. The multi-color nature of these features demonstrates a sharp velocity gradient over just several adjacent spectral channels with linewidths comparable to CO(2$\rightarrow$1) emission observed in individual molecular clouds in other nearby external galaxies as reported by the PHANGS-ALMA survey \citep{leroy2021}. In our own Galaxy, the atomic-to-molecular ($\hi$-to-H$_2$) phase transition in the dense regions of GMCs, arising from the combined effect of dust attenuation of far-ultraviolet radiation and H$_2$ self-shielding (\citealt{bialy2015} and references therein), traces sites of star formation. \citet{glover2012} argue that the transition from the WNM to CNM phase is an important step towards star formation. Studying the gas transition from $\hi$ to H$_2$-dominated regions that fuel star formation within the Bar and comparing with observational studies of Milky Way GMCs (e.g., \citealt{young-lee_2012, burkhart2015, wang2020}) will be an invaluable step towards a complete theory of star formation.

Towards blueshifted velocities (e.g., 131 km s$^{-1}$), a prominent population of compact, spatially anomalous $\hi$ features surround the SMC. Considering the general head-tail morphology of associated molecular gas by \citet{diTeodoro2019}, a majority of these anomalous features likely originate from strong star-formation-driven outflows \citep{McClure-Griffiths2018}, though the uncertain geometry of the SMC conspires against a definitive measure of the escaping gas. Nevertheless, these novel observations will inform the comparison between kinematic and geometric properties of infall/outflow models. 

\subsection{$\hi$ Turbulent Properties}\label{subsec:turb_properties}
The extension of a single component power-law down to physical scales three times smaller than previously measured demonstrates processes such as star formation, which injects energy on several hundred pc scales \citep{MacLow2004}, do not produce any special feature or deviation in the large-scale SPS of the SMC. Either stellar feedback is not the dominant source of energy injection at these scales and large-scale driving is more dominant (at least within the SMC; \citealt{szotkowski2019}), or energy is being injected over a wide range of scales without preference at any particular scale (e.g., \citealt{norman1996}). This is also congruous with the notion that the power-law form of the power spectrum arises from the fractal geometry of the ISM; that is, the ISM displays hierarchical structuring over the entire range of spatial scales \citep{stanimirovic1999, elmegreen1996}. Interestingly, the average spectral index measured for all slopes over velocity range of 90 km s$^{-1}$ to 190 km s$^{-1}$ is $\gamma$ = $-$2.85$\pm$0.07, which is slightly shallower than $\gamma$ = $-$3.04$\pm$0.04 measured by \citet{stanimirovic1999}. Shallower slopes across the entire velocity range indicate more power on smaller scales, possibly from more abundant cold $\hi$ that is resolved by our higher angular resolution GASKAP-HI observations. On the other hand, the differences in the average slopes could also be driven from a statistical bias (e.g., small number statistics for the large-scale power). Following the application of the discussion from \citet{falconer1997} and \citet{stanimirovic1999}, we exploit the Fourier Transform relationship between the autocorrelation function and power spectrum to connect $\gamma$ to the fractal function with the relation
\begin{equation}
    -\gamma-2=4-2s, 
\end{equation}
where $s$ is the box-counting dimension. This value quantifies the fractal nature of a subset of three-dimensional Euclidean space, R$^3$, and provides an upper limit for the fractal dimension (also known as the Hausdorff dimension). For example, a perfectly smooth square has a fractal dimension of 2, while an object with a non-integer fractal dimension indicates self-similarity over a range of scales and thus the presence of hierarchical structure. We determine the $s$ of the combined ASKAP and Parkes images of $\hi$ emission to be 1.58$\pm$0.04, in agreement with the fractal analysis in \citet{stanimirovic1999}. The fractal dimension observed in molecular clouds is 1.36$\pm$0.02 from scales ranging from 100 pc to 0.02 pc \citep{falgarone1991}. This statistically significant difference indicates separate mechanisms drive the dynamics in the WNM and molecular clouds.

Analogously, a clustering analysis applied to the upper main-sequence stars in the SMC shown by \citet{sun2018} shows young stellar structures, much like the gas, are also organized in a hierarchical fashion. The consistent description for both the gas and stars indicates similar mechanisms, such as turbulence and/or hierarchical gravitational fragmentation of the denser gas that forms stars \citep{vazquez2019}, influence the properties of the $\hi$ across spatial scales extending from kpc down to the scales of individual molecular clouds on sub-parsec scales. We note, however, that a comprehensive analysis of the fractal structure, especially at individual spectral channels, should account for the effects of self-absorption and difficulty in disentangling the relative contributions of the CNM and WNM to the overall $\hi$ emission.

\subsection{Properties of an Anomalous $\hi$ Feature}\label{subsec:hvc}

\begin{figure*}
\centering{
\includegraphics[width=\textwidth]{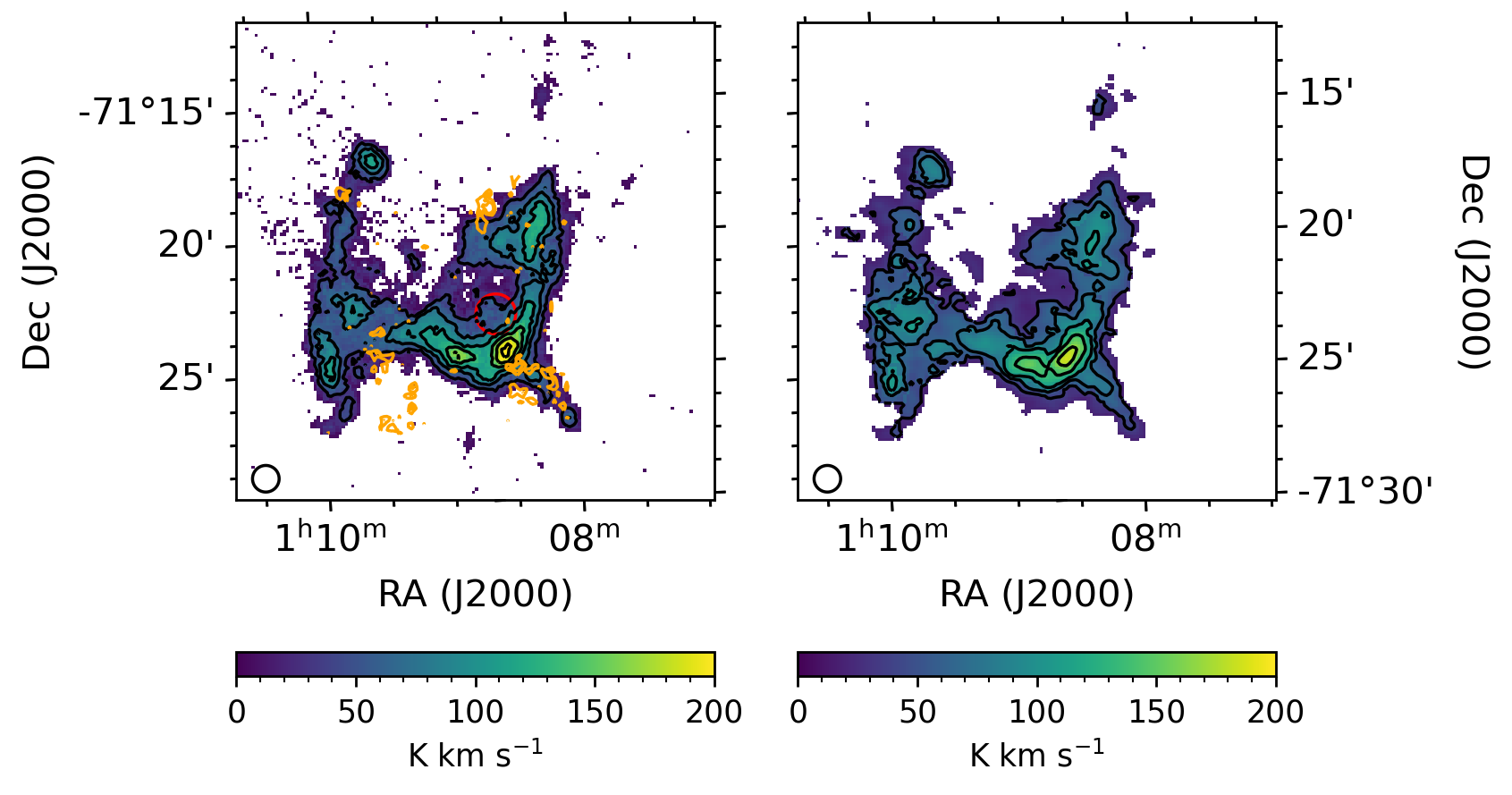}
\caption{\label{fig:hvc_mom0} Combined ASKAP+Parkes $\hi$ masked integrated intensity images of an HVC centered on RA = 01$^{\rm h}$08$^{\rm m}$45$^{\rm s}$, Dec = $-$71$^{d}$20$^{\rm m}$55$^{\rm s}$ produced by the source finding algorithm in {\tt BBarolo}. The projected physical extent of this region is 300 pc at a distance of 60 kpc. The image made from our improved pilot survey is on the left and the image from the test observations \citep{McClure-Griffiths2018} is on the right. In both cases, the detected emission was integrated over $90 \leq $v$_{\rm LSRK} \leq 118$ km s$^{-1}$. The contour levels increase in steps of 29.5 K km s$^{-1}$ and range from 5 K km s$^{-1}$ to 300 km s$^{-1}$, inclusive. The red circle with a radius corresponding to 13 pc denotes the approximate size of a shell-like feature identified in position-velocity space. The orange contours in the left panel show the CO integrated intensities at 0.02 and 0.04 K km s$^{-1}$ from \citet{diTeodoro2019}. The size of the restoring beam for the ASKAP data sets is denoted by the circle in the lower left corner.}}
\end{figure*} 

The test observations presented in \citet{McClure-Griffiths2018} revealed a considerable population of spatially and kinematically anomalous $\hi$ features that extend upwards of 2 kpc from the main body of the SMC. The increase in spectral resolution and overall sensitivity of our pilot survey observations facilitate more robust constraints of the physical properties of these Magellanic HVCs --- such as their mass, kinematics, and morphology. These in turn trace fundamental properties of the circumgalactic environment around the SMC and Milky Way halo like the total amount of outflowing gas and halo pressure by relating head-tail morphologies with ram pressure and/or instabilities.

We illustrate the power of our new pilot data to trace anomalous HVCs in the Magellanic System by focusing on an outflowing HVC centered on RA = 01$^{\rm h}$08$^{\rm m}$45$^{\rm s}$, Dec = $-$71$^{d}$20$^{\rm m}$55$^{\rm s}$ that entrains several compact $^{12}$CO(2$\rightarrow$1) clouds \citep{diTeodoro2019}. The left and right panels of Figure~\ref{fig:hvc_mom0} compare the integrated intensity images of this HVC from the pilot and test observations, respectively. We utilize the {\tt SEARCH} task in {\tt BBarolo}, a software package for fitting 3D tilted-ring models to emission-line cubes \citep{diTeodoro2015}, in order to exclude noisy pixels from the integration of the signal. We set the SNR threshold for genuine signal to any pixel with SNR $>$ 5 over the span of two or more adjacent channels, applied within the LSRK velocity range of 95 km s$^{-1}$ to 119 km s$^{-1}$. The remaining task parameters use default values. The images from the pilot and test observations show similar large-scale morphology. However, the small-scale structures in the pilot data displays more pronounced density gradients in individual clumps. This is due to the twofold increase in sensitivity, fourfold increase in spectral resolution (facilitating more narrow-line features to be resolved), and more complete sampling of higher spatial frequencies from the additional antennas at longer baselines. For example, let us focus on the northeast condensation with a peak at RA = 01$^{\rm h}$09$^{\rm m}$35$^{\rm s}$, Dec = $-$71$^{d}$17$^{\rm m}$15$^{\rm s}$. The mean column densities are roughly similar at 1.3$\times$10$^{20}$ cm$^{-2}$ and 8.2$\times$10$^{19}$ cm$^{-2}$ in the pilot and test images, respectively, giving a volume density $n\sim$ 1.6 cm$^{-3}$ and 1.0 cm$^{-3}$ assuming spherical symmetry, an angular extent of 90$''$, and a distance of 60 kpc. 

The full potential for constraining the physical properties of these HVCs is realized through the increased spectral resolution. For example, we break up a $2.5'\times2.5'$ region centered on the peak intensity of this condensation to the northeast for the pilot and test data into a 4$\times$4 grid to independently sample the emission across 16 restoring beam elements and compute the integrated spectrum for each sample. We then shift each integrated spectrum to the same peak velocity and average the profile. Fitting a single Gaussian function to these averaged profiles returns $v_{\rm FWHM}$ of 6.9$\pm$0.5 km s$^{-1}$ and 13$\pm$2 km s$^{-1}$ for the pilot and test data, respectively; the differences arising from the differences in spectral resolution. These $v_{\rm FWHM}$ provides an upper limit for the kinetic temperature of the gas, $T_{\rm max, k}$ = 21.866$\times v_{\rm FWHM}^2$ of 1000$\pm$200 K and 4000$\pm$1000 K. We can use the $T_{\rm max, k}$ and $n$ to obtain an estimate of the pressure of this condensation through the approximation of an ideal gas, $P/k = nT_{\rm max, k}$, where $P$ is the pressure and $k$ is the Boltzmann constant. The resulting upper limit pressures for the pilot and test data are respectively $P/k = 1600\pm300$ K cm$^{-3}$ and $P/k = 4000\pm1000$ K cm$^{-3}$. The respective decrease of nearly 60\% and 70\% between the upper limit pressures and final uncertainties demonstrates that the superior spectral and angular resolution of the pilot data is crucial for constraining important physical properties. This also demonstrates how increased angular and spectral resolutions work together to improve estimates of intrinsic line widths.

\begin{figure*}
\centering{
\includegraphics[width=\textwidth]{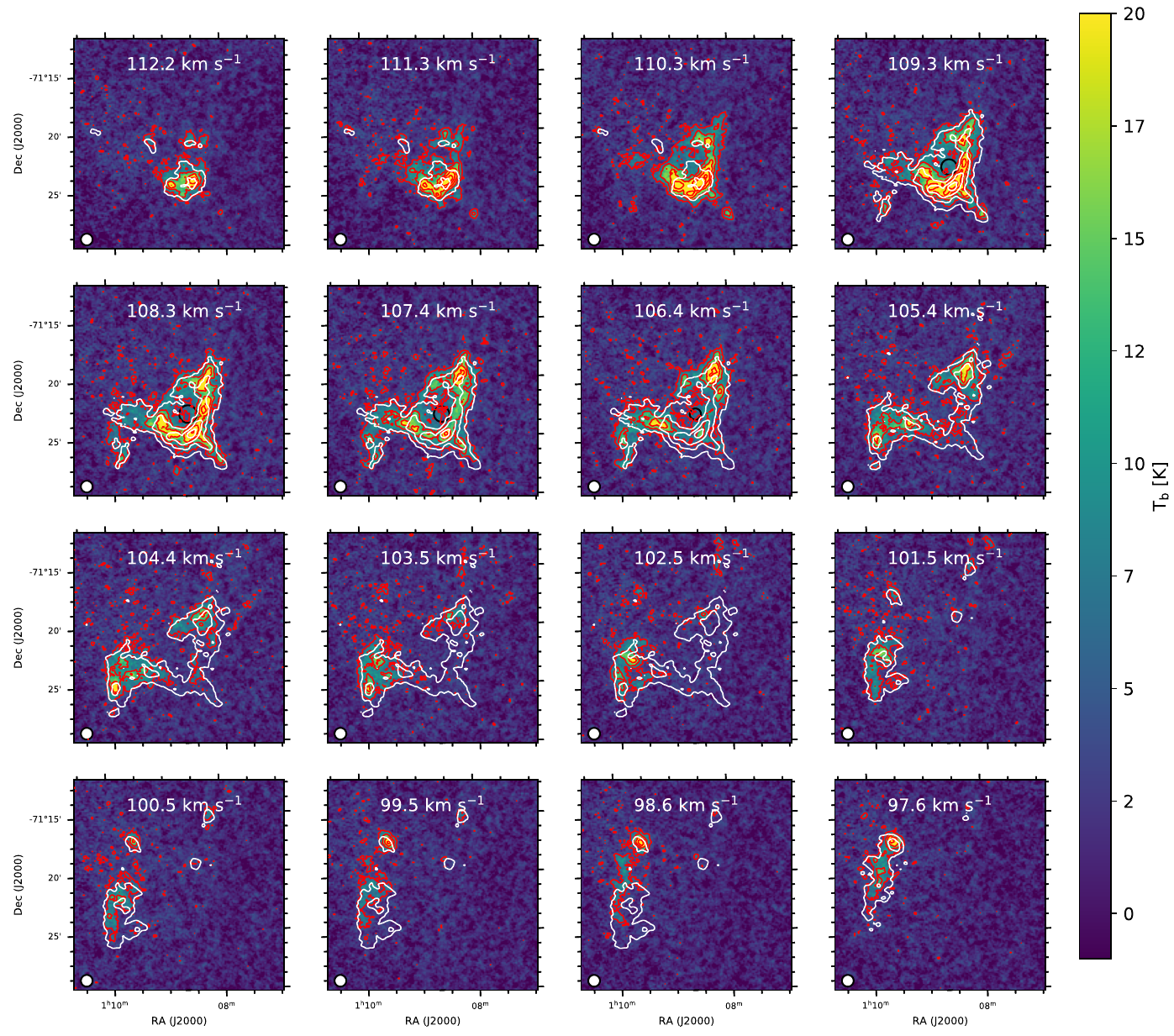}
\caption{\label{fig:hvc_chanmaps} Combined ASKAP+Parkes $\hi$ channel maps of an HVC centered on RA = 01$^{\rm h}$08$^{\rm m}$45$^{\rm s}$, Dec = $-$71$^{d}$20$^{\rm m}$55$^{\rm s}$. The black contours denote emission from the pilot cube, while the white contours trace emission from the lower spectral resolution test cube. Both sets of contours range between 5 K and 30 K and increase in increments of 6.25 K. The increased spectral resolution of the pilot cube resolves several notable features including a $\hi$ hole between 110.3 and 105.4 km s$^{-1}$. The approximate projected size of the shell at these velocities is denoted by a black circle, assuming uniform expansion and spherical symmetry. The restoring beam of the pilot data is shown in the lower left corner of each panel.}}
\end{figure*}

\begin{figure*}
\centering{
\includegraphics[height=0.8\textheight]{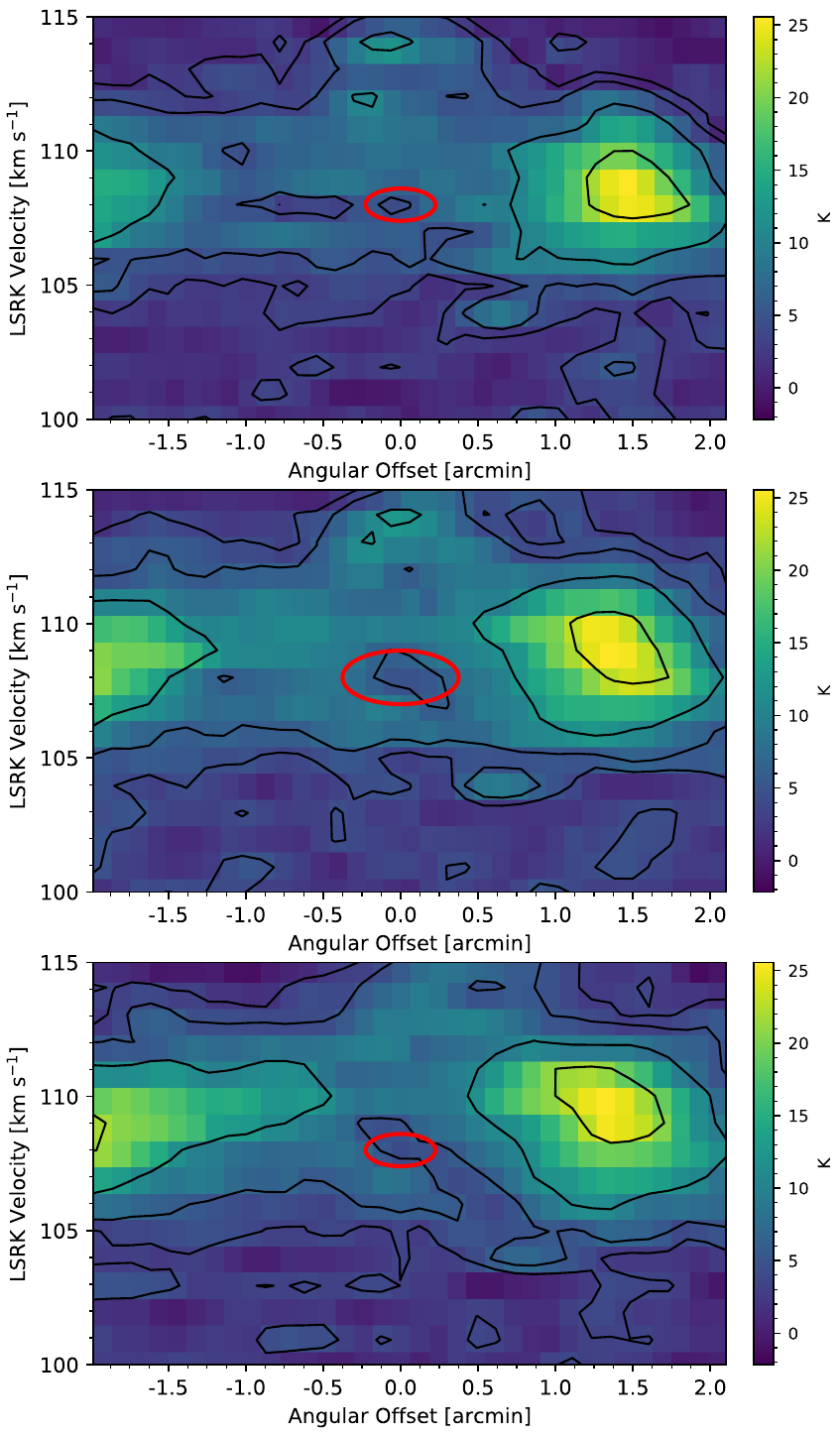}
\caption{\label{fig:shell_pv} Combined ASKAP+Parkes $\hi$ position-velocity distribution for three slices along RA. The top panel represents a slice 30$''$ above the center of the circle in the left panel of Figure~\ref{fig:hvc_mom0} that intercepts the $\hi$ shell at an angle of 50$^{\circ}$ from a horizontal slice through the center located at RA = 01$^{\rm h}$08$^{\rm m}$38$^{\rm s}$, Dec = $-$71$^{d}$23$^{\rm m}$00$^{\rm s}$. The middle panel represents a slice directly through the center, and bottom panel shows a slice 30$''$ below the center that intercepts the shell at a $-$50$^{\circ}$ angle below a horizontal slice through the center. Assuming the geometry outlined for a uniformly expanding shell with $v_{\rm exp}$ = 2.0 km s$^{-1}$ (much smaller than the typical thermal line widths expected for the WNM), we show the predicted position-velocity structure as a red ellipse. The emission contours are at levels of 3, 5 K, 10 K, and 20 K.}}
\end{figure*}

The CO integrated intensity contours from the observations of \citet{diTeodoro2019} overlap well both spatially and spectrally (see their Figure 3) with the southern edge of the most pronounced $\hi$ density gradient in this cloud centered on RA = 01$^{\rm h}$08$^{\rm m}$33$^{\rm s}$, Dec = $-$71$^{d}$24$^{\rm m}$21$^{\rm s}$. The emission distribution in individual spectral channels in Figure~\ref{fig:hvc_chanmaps} reveal a shell-like structure appearing first at a velocity of 110.3 km s$^{-1}$ before completely breaking up at 103.4 km s$^{-1}$ along the northern edge of this density gradient.

The distribution of emission in position-velocity space of several RA cuts in Figure~\ref{fig:shell_pv} hints at a characteristic ellipse structure expected from an expanding shell, indicating spherical symmetry in 3D space. We determine the center of the apparent shell-like feature to be RA = 01$^{\rm h}$08$^{\rm m}$38$^{\rm s}$, Dec = $-$71$^{d}$23$^{\rm m}$00$^{\rm s}$ and measure $v_{\rm min}$ = 106$\pm$1 km s$^{-1}$ and $v_{\rm max}$ = 110$\pm$1 km s$^{-1}$ and thus a systemic velocity $v_{\rm sys}$ = 108$\pm$1 km s$^{-1}$. Assuming a uniformly expanding shell, the projected sizes in position-velocity space are respectively
\begin{equation}\label{eq:proj_v}
    v_{\rm proj} = v_{\rm exp}\cos{\theta_{\rm cut}}+v_{\rm sys}
\end{equation}
and
\begin{equation}\label{eq:proj_r}
    R_{\rm pos} = R_{\rm ang}\cos{\theta_{\rm cut}},
\end{equation}
where $\theta_{\rm cut}$ represents the angle at which a horizontal cut is above and below another horizontal cut through the center of the feature in position-position space and $R_{\rm ang}$ is the angular projected radius at the center and is determined to be 13 pc at the distance of adopted distance of 60 kpc. The line-of-sight (LOS) geometry is estimated using 
\begin{equation}\label{eq:phi}
    \phi = \cos^{-1}\left(\frac{v_{\rm LSRK} - v_{\rm sys}}{v_{\rm exp}}\right)
\end{equation}
\begin{equation}\label{eq:pos_r}
    R_{\rm proj}=R_{\rm ang}\sin{\phi}, 
\end{equation}
where $v_{\rm LSRK}$ is again the observed radial velocity in the LSRK reference frame. Using these $v_{\rm min}$ and $v_{\rm max}$ and measuring the velocity of the emission at the location of a horizontal cut that intercepts the shell at an angles of $\theta_{\rm cut}=50^{\circ}$ and $\theta_{\rm cut}=-50^{\circ}$ above and below the center, we estimate the expansion velocity $v_{\rm exp}$ to be 2$\pm$1 km s$^{-1}$. Using Equations~\ref{eq:proj_v}-\ref{eq:pos_r}, we show the projected size in the relevant channel maps and in position-velocity space. 

A $v_{\rm exp}$ of 2$\pm$1 km s$^{-1}$ is roughly five times smaller than typical thermal line widths for the WNM, making it difficult to discern likely origin scenarios based on input energy estimates for possible progenitors. A search for bright ($G$ $<$ 20), blue (color $-$0.5 $<$ BP$-$RP $<$ 0) stars with parallaxes corresponding to distances larger than 55 kpc within the Gaia Early Data Release 3 catalogue \citep{gaiaMission, GaiaDR3} shows dozens of stars in the field. However, we cannot point to an isolated population as the potential progenitor of this expanding shell without further filtering based on radial velocity measurements, which are currently unavailable. Moreover, the measured mass within the aperture encompassing the feature is $\sim$2240 \msun, implying a corresponding thermal energy of 8.9$\times$10$^{47}$ erg. This is much lower than the input from a single core-collapse supernova explosion \citep{hartmann1999}. The absence of an isolated stellar association, small $v_{\rm exp}$ relative to the thermal linewidth of the WNM, and correspondingly low thermal energy point to a purely turbulent origin. \citet{daigle2007} note the presence of structures at similar scales as this feature in simulated $\hi$ image cubes with pure turbulent motions. Regardless of the true origin, this investigation demonstrates that our GASKAP-HI data provide a novel view into the kinematic properties of the diverse population of exo-galactic clouds surrounding the SMC.

\begin{figure*}
\centering{
\includegraphics[width=\textwidth]{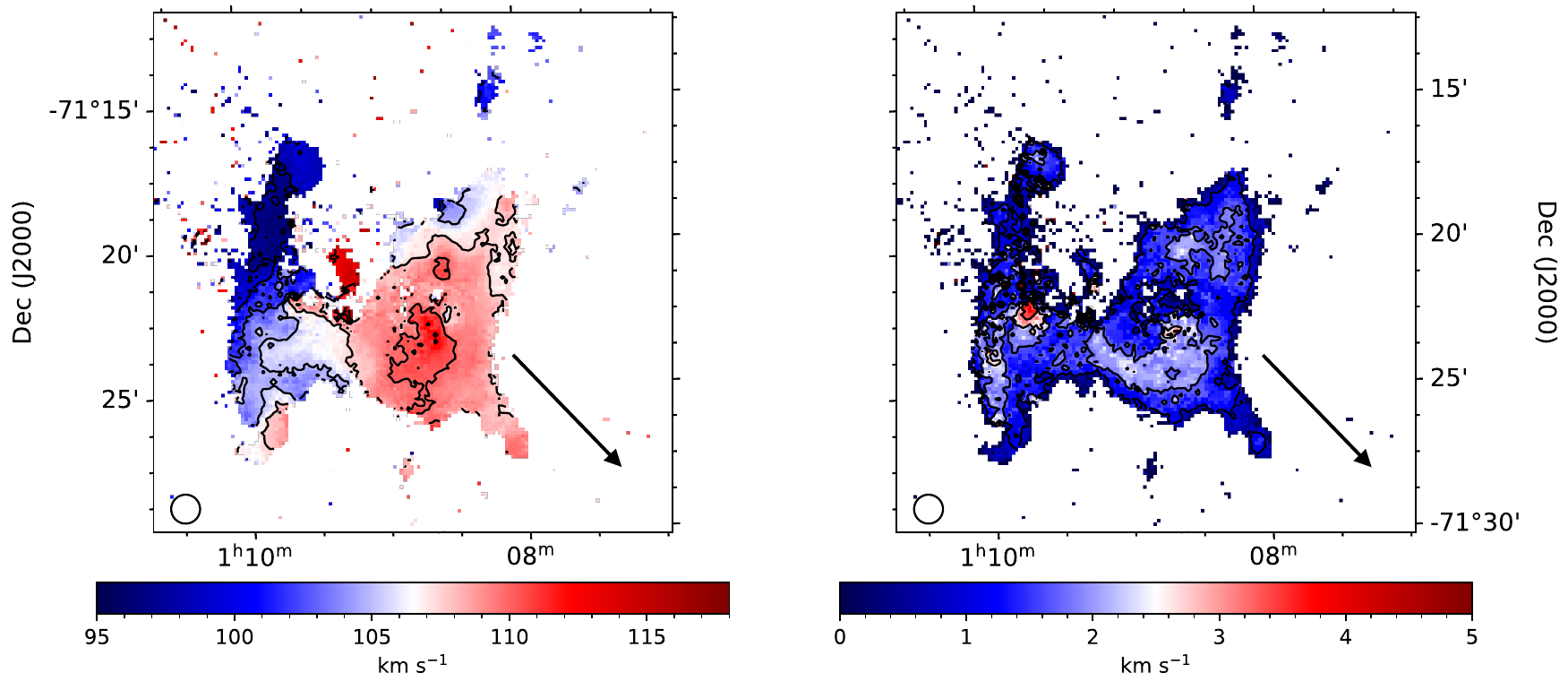}
\caption{\label{fig:hvc_mom1_2} Combined ASKAP+Parkes $\hi$ maps of the intensity-weighted velocity field (left) and velocity dispersion (right) of an HVC centered on RA = 01$^{\rm h}$08$^{\rm m}$45$^{\rm s}$, Dec = $-$71$^{d}$20$^{\rm m}$55$^{\rm s}$ produced by the source finding algorithm in {\tt BBarolo}. The velocity field contours range inclusively between the levels of 92.5 km s$^{-1}$ and 115 km s$^{-1}$ in steps of 2.5 km s$^{-1}$. The contours of velocity dispersion begin at a level of 1 km s$^{-1}$ and increase up to 4 km s$^{-1}$ in steps of 0.75 km s$^{-1}$. The gradient in the velocity field across the entirety of the cloud and complexity of the velocity dispersion contours --- especially towards the northwest spur --- are qualitatively consistent with this HVC being a wind-swept cloud from a Galactic wind emanating from the SMC. The arrow denotes the direction towards the nearest prominent star formation region, NGC 371/395. The size of the restoring beam is shown in the lower left corner of both panels.}}
\end{figure*}

The presence of galactic-scale external winds can explain the global observed morphological and kinematic properties of this cloud. Galactic winds shape the structures of the ISM, altering the kinematic and morphological properties of the gas. Hot winds interacting with cold clumps of dense gas can excite molecular species \citep{pon2012, girichidis2021}, transfer mass, momentum, and energy \citep{vijayan2020, scheider2020}, and strip the outer surface layer, forming elongated, filamentary structures that trail the main body of the cloud (e.g., \citealt{banda-barragan2016, gronke2021}). \citet{diTeodoro2019} suggests a scenario in which this particular complex is launched from NGC 371/395, the nearest prominent star formation region located south-west. The intensity-weighted velocity field shown in the left panel of Figure~\ref{fig:hvc_mom1_2} shows a clear gradient across the entirety of the cloud. Such velocity gradients in discrete clouds are often interpreted as rotation. However, the closed isovelocity contours in the western side of the cloud that overlap with the expanding bubble, absence of a clear major axis, and disturbed structure in the velocity dispersion map instead indicate these motions are more likely due to shear. The population of discrete cloudlets to the northwest in the individual spectral channels and moment maps is also consistent with the expectation of the formation of head-tail structures from the stripping of outer gas layers seen in simulations of wind-cloud interactions and observations \citep{phillips2009, gronnow2017, gronnow2018,banda-barragan2019, banda-barragan2021}. 

\begin{figure}
\centering{
\includegraphics[width=\columnwidth]{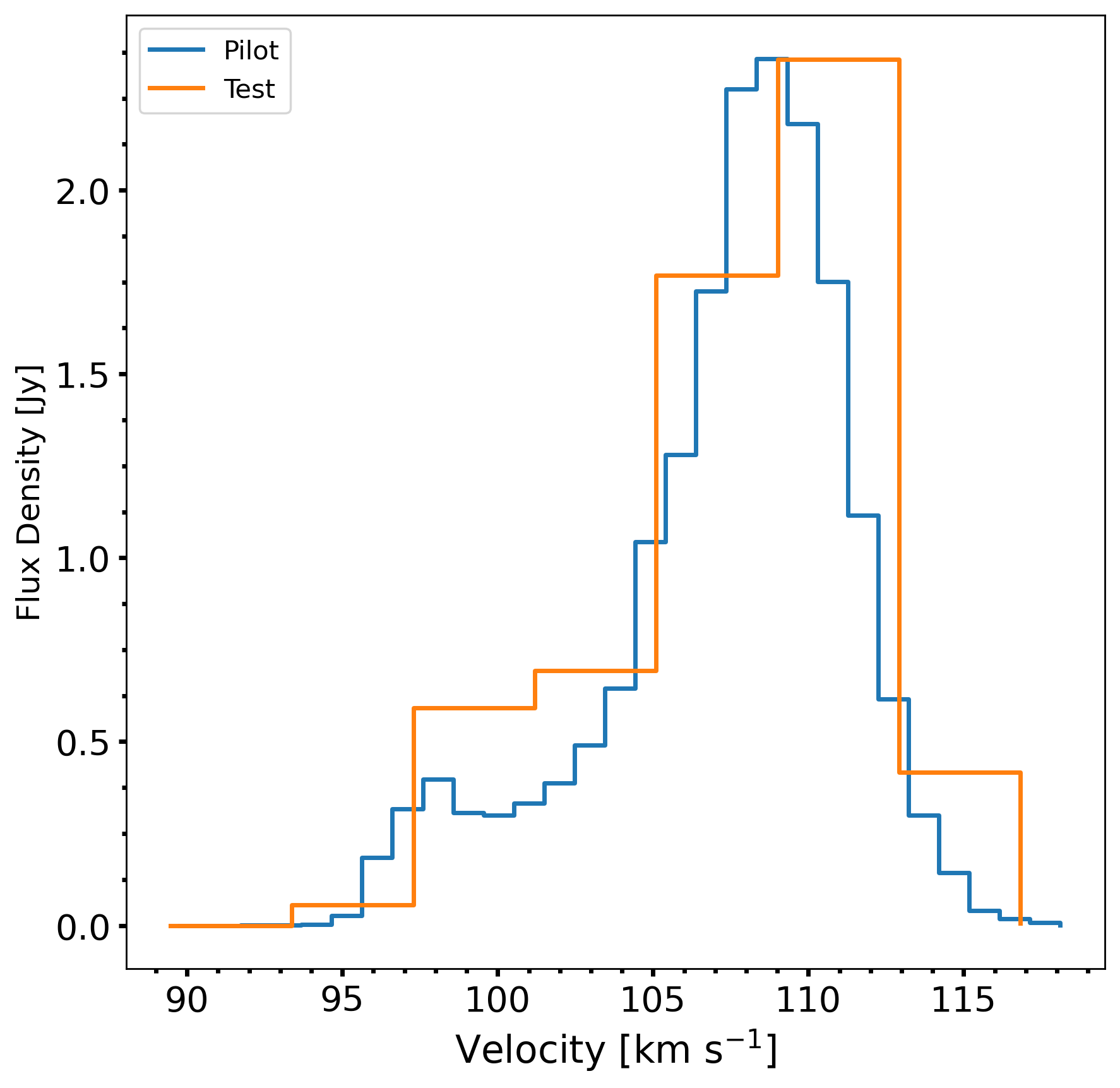}
\caption{Combined ASKAP+Parkes $\hi$ integrated spectrum of the HVC measured over the area shown in Figures~\ref{fig:hvc_mom0} and \ref{fig:hvc_chanmaps}.
\label{fig:hvc_spectrum}}}
\end{figure}

The integrated spectrum shown in Figure~\ref{fig:hvc_spectrum}, measured over the field of view shown in Figures~\ref{fig:hvc_mom0} and~\ref{fig:hvc_chanmaps}, further demonstrates the advantage of the increased spectral resolution in deriving physical properties of HVCs around the SMC. The pilot flux profile shows the multi-component kinematic properties of this cloud, which is only hinted at in the coarser spectrum. Inspection of the individual channel maps for velocities 105 km s$^{-1}$ to 97 km s$^{-1}$ shows the flux offset between the two profiles occurs because a significant amount of emission happens to sit at the boundary of a coarser spectral channel in the test data, thus artificially extending the velocity range of the emission distribution. Again assuming a distance of 60 kpc \citep{McClure-Griffiths2018}, we achieve a more reliable measure of the $\hi$ mass to be 15,100$\pm300$ \msun, as opposed to an overestimated value of 19,600$\pm$100 \msun. This discrepancy is likely related to improvements made in the calibration and processing pipeline between November 2017 and December 2019. 

These data enable us to derive a diverse range of physical properties for this single HVC. Future detailed cataloging and study of the HVC population around the SMC will inform mass outflow/inflow models to help determine whether such features are outflowing or accreting, aid in constraining the overall dynamics of the SMC, and trace the pressure equilibrium between cold and warm $\hi$ phases through Gaussian decomposition algorithms, such as the Regularized Optimization for Hyper-Spectral Analysis (ROHSA) package \citep{marchal2019} or GaussPy+ \citep{lindner2015, reiner2019}. 

\section{Summary}\label{sec:summary}
Utilizing a combination of ASKAP and Parkes GASS $\hi$ data to obtain a cube covering all angular scales, we presented the most sensitive and detailed $\hi$ images of the SMC ever achieved. These images were produced by our custom imaging pipeline, which is designed to optimize a joint deconvolution approach in a cluster environment in order to recover an accurate representation of the extended sky brightness distribution. In addition to a full description of the calibration procedures, imaging pipeline, and quality assessments, we demonstrated that:
\begin{itemize}
  \item the spatial power spectrum is a powerful tool in determining the correct scaling factor applied to the total power observations when combining interferometric and single dish data. For example, an overestimated scaling factor results in a \emph{decrease} in power at small scales because the combination effectively washes out the small-scale structure present in an interferometer-only image. 
  \item The final SMC image cube possesses a rms noise level of 1.1 K per 0.98 km s$^{-1}$ spectral channel and an angular resolution of 30$''$ (8 pc at the distance of the SMC). The noise properties of the final data are generally consistent across the FoV until the sensitivity begins to drop at the edge of the PAF footprint. A low-level grid pattern is apparent, likely stemming from numerical artifacts from the Fourier Transforms that invert the gridded visibilities to the image domain during major iterations. The noise scales as predicted by \citet{dickey2013} with the resolution of the restoring beam, indicating these GASKAP-HI data are a breakthrough in terms of the compromise between rms brightness temperature and angular resolution.
  \item Our power spectrum analysis shows the distribution of $\hi$ emission in the SMC is well described by a single component power-law, consistent with the interpretation that the ISM is fractal in nature and that stellar feedback is probably not highlighted at any particular scale. Stochastic processes, such as turbulence, therefore influence the structure of the ISM from scales ranging from kpc down to sub-parsecs. 
  \item A comparison of the basic properties of an outflowing $\hi$ complex derived from test data with $\sim$4$\times$ worse spectral resolution relative to these new pilot observations demonstrates the power of our pilot data to constrain important ISM properties. The increase in spectral resolution allows us to resolve an expanding $\hi$ hole. The low thermal energy and expansion velocity suggests a purely turbulent origin for this feature. 
\end{itemize}

These new $\hi$ images of the SMC show spectacular details on scales from 5 kpc extending down to 10 pc and reveal previously unresolved narrow line features. A separate paper will present the abundance of absorption detections in this same field. These pilot data facilitate the study of how $\hi$ contributes to the conditions necessary for star formation, probe the hierarchical turbulent spectrum down to physical scales three times smaller than previously measured, and truly characterize the multi-phase make up of the interstellar medium. GASKAP-HI provides an unequivocal view of the nearby $\hi$ associated with the detailed structures observed in the dust and molecular gas. Additionally, the lessons learned from imaging extremely extended emission for the GASKAP-HI survey will inform the data processing of future ultra wide-field surveys to be undertaken by the Square Kilometer Array that will face identical imaging challenges. The increase in brightness temperature sensitivity, coupled with the unparalleled angular and spectral resolution, ensures the GASKAP-HI survey will provide touchstone data products for the study of $\hi$ in the  Milky Way and Magellanic System for the next decade and beyond.

\begin{acknowledgements}
We thank the anonymous expert reviewer whose comments helped to focus the presentation and discussion of these data products. The Australian SKA Pathfinder is part of the Australia Telescope National Facility which is managed by CSIRO. Operation of ASKAP is funded by the Australian Government with support from the National Collaborative Research Infrastructure Strategy. ASKAP uses the resources of the Pawsey Supercomputing Centre. Establishment of ASKAP, the Murchison Radio-astronomy Observatory and the Pawsey Supercomputing Centre are initiatives of the Australian Government, with support from the Government of Western Australia and the Science and Industry Endowment Fund. We acknowledge the Wajarri Yamatji people as the traditional owners of the Observatory site. Pipeline development was tested on the OzSTAR supercomputer under the project code, oz145, which is available through Swinburne University's Centre of Astrophysics and Supercomputing.

This research was supported by the Australian Research Council (ARC) through grant DP190101571.  N.M.-G. acknowledges the support of the ARC through Future Fellowship FT150100024. G.~A. and J.~F.~G. acknowledge support from the State Agency for
Research (10.13039/501100011033) of the Spanish MCIU, through grants
AYA2017-84390-C2-1-R and PID2020-114461GB-I00 (co-funded by FEDER) and the ``Center of Excellence Severo Ochoa'' award
for the Instituto de Astrof\'{\i}sica de Andaluc\'{\i}a (SEV-2017-0709). C.~E.~M.~is supported by an NSF Astronomy and Astrophysics Postdoctoral Fellowship under award AST-1801471.

\end{acknowledgements}

\bibliographystyle{pasa-mnras}
\bibliography{refs}

\end{document}